\documentclass[aps,prd,superscriptaddress,11pt,showpacs,notitlepage]{revtex4}
	\usepackage{graphicx}
	\usepackage{amsmath,amssymb,mathrsfs}
\usepackage{epsf}
\usepackage{epsfig}

\usepackage{amsmath}

\usepackage{amssymb}

\numberwithin{equation}{section}

\newcommand{\be}{\begin{equation}}
\newcommand{\ee}{\end{equation}}
\newcommand{\bea}{\begin{eqnarray}}
\newcommand{\eea}{\end{eqnarray}}

\newcommand{\bb}{\bibitem}
 
\newcommand{\eqn}{\begin{eqnarray}}
\newcommand{\eqnx}{\end{eqnarray}}

\begin{document}
\title{Magnetothermodynamics of BPS baby skyrmions}
\author{C. Adam}
\affiliation{Departamento de F\'isica de Part\'iculas, Universidad de Santiago de Compostela and Instituto Galego de F\'isica de Altas Enerxias (IGFAE) E-15782 Santiago de Compostela, Spain}
%\author{C. Naya}
%\affiliation{Departamento de F\'isica de Part\'iculas, Universidad de Santiago de Compostela and Instituto Galego de F\'isica de Altas Enerxias (IGFAE) E-15782 Santiago de Compostela, Spain}
\author{T. Romanczukiewicz}
\affiliation{Institute of Physics,  Jagiellonian University,
Reymonta 4, Krak\'{o}w, Poland}
\author{J. Sanchez-Guillen}
\affiliation{Departamento de F\'isica de Part\'iculas, Universidad de Santiago de Compostela and Instituto Galego de F\'isica de Altas Enerxias (IGFAE) E-15782 Santiago de Compostela, Spain}
\author{A. Wereszczynski}
\affiliation{Institute of Physics,  Jagiellonian University,
Reymonta 4, Krak\'{o}w, Poland}
%%\author{W. J. Zakrzewski}
%%\affiliation{Department of Mathematical Sciences, University of Durham, Durham DH1 3LE, U.K.}

\begin{abstract}
The magnetothermodynamics of skyrmion type matter described by the gauged BPS baby Skyrme model at zero temperature is investigated. We prove that the BPS property of the model is preserved also for boundary conditions corresponding to an asymptotically constant magnetic field. The BPS bound and the corresponding BPS equations saturating the bound are found. Further, we show that one may introduce pressure in the gauged model by a redefinition of the superpotential. Interestingly, this is related to non-extremal type solutions in the so-called fake supersymmetry method. Finally, we compute the equation of state of magnetized BSP baby skyrmions inserted into an external constant magnetic field $H$ and under external pressure $P$, i.e., $V=V(P,H)$, where $V$ is the "volume" (area) occupied by the skyrmions. We show that the BPS baby skyrmions form a ferromagnetic medium. 
\end{abstract}
\maketitle 
%%%%%%%%%%%%%%%%%%%%%%%%%%%%%%%%%%%%%%%%%
\section{Introduction}
%%%%%%%%%%%%%%%%%%%%%%%%%%%%%%%%%%%%%%%%%
The Skyrme model \cite{skyrme} is considered one of the best candidates for an effective model of low energy QCD. Using results from the large $N_c$ expansion, it is known that the proper degrees of freedom in this limit are mesons, while baryons emerge as collective excitations, i.e., solitons called skyrmions, with an identification between baryon number and topological charge. To get phenomenologically precise relations between solitons and baryons (nuclei), one has to perform the standard semiclassical quantization of the spin and isospin degrees of freedom, as well as include the electromagnetic interaction, which obviously contributes to the masses of particles. Fortunately, although the Skyrme model has not  yet been derived from the underlying microscopic quantum field theory, its coupling to the electromagnetic field is completely determined by the symmetries and anomalies of QCD \cite{wit1}.  The resulting $U(1)$ gauged Skyrme model is rather difficult to analyse, and the electromagnetic properties of nucleons as well as atomic nuclei, although very important, could not yet be extracted in the full nonlinear Skyrme-Maxwell description. The electric part of the energy of the nuclei is typically approximated by the Coulomb energy \cite{Coul}, where the back reaction of the Maxwell field on the Skyrme matter field is not taken into account. 
Let us remark that some first numerical results for the Skyrme model minimally coupled to the electromagnetic field (but without the anomalous or Wess-Zumino-Witten term contribution) have been found in \cite{GSk}. Further, very recently some knotted soliton solutions have been obtained for the $S^2$ restriction of the minimally gauged Skyrme model i.e., the gauged Faddeev-Skyrme-Niemi model, however within the toroidal ansatz which limits the solutions to the charge $Q=1,2$ sectors \cite{shnH}.   	
\\
As has been mentioned already, a precise derivation of the Skyrme model (or in fact any effective low energy model) from QCD is one of the most urgent, however still unsolved, tasks in modern theoretical physics. The lack of a systematic derivation means that the precise form of the Skyrme type action is not known. The usual assumption (based on a perturbative approach) restricts the model to three terms: the sigma model part (Dirichlet energy), the Skyrme term (obligatory to avoid the Derrick arguments for the non-existence of static solutions) and a potential (providing a mass for the perturbative pionic excitations). It is, however, one of the main problems of the usual Skyrme model that it leads to unphysical binding energies, which are in strong disagreement with the experimental data. The underlying reason for this is that the usual Skyrme model is not a BPS theory, i.e., the energies of skyrmions are not linearly related to their topological charges.  As atomic nuclei seem to be close to BPS objects (the masses are almost linear in the baryon charge with a $1\%$ deviation, at most), the corresponding effective model should be a (near) BPS one. There exist two quite different realizations of this concept. The first proposal is based on the observation that the inclusion of infinitely many vector mesons (Kaluza-Klein modes) can bring the original Skyrme model towards the $(4+0)$ Yang-Mills action \cite{Sut1}, \cite{hidden}. In the second proposal, the crucial observation is that within all Skyrme type Lagrangians (i.e., with no additional fields) there exists a special one with the BPS property. It has a rather simple form and consists of two mutually balancing terms: a derivative part (the baryon  (topological) current squared) and a potential \cite{BPS-Sk}.  Moreover, this model possesses the volume preserving diffeomorphism symmetry, which allows to interpret it as a field theoretical description of the liquid droplet model. In addition, the static energy-momentum tensor of the model is the energy-momentum tensor of a perfect fluid, further strengthening the case for this interpretation. As a consequence, there are infinitely many solitonic solutions saturating a topological bound, which leads to a linear energy - topological charge relation. Therefore, the classical binding energies are zero. Further, finite binding energies have been recently derived by taking into account the semiclassical quantization of the spin-isospin degrees of freedom, the Coulomb interaction as well as the isospin breaking potential. The obtained values are in very good agreement with the nuclear data and the semi-empirical (Weizs\"acker) formula, especially for higher nuclei \cite{BPS nucl}, \cite{Marl}. This result allows to consider the BPS Skyrme model as a serious candidate for a lowest order approximation of the correct effective model of QCD at low energies, especially for the bulk quantities. 
\\
In addition to the binding energies, there are many properties of nuclei and nuclear matter which should be understood within the framework of the (near) BPS Skyrme model. It is another advantage of this model that, due to its generalized integrability and BPS nature (solvability), many relevant questions can be answered in an analytical manner. One of the most important ones is related to the thermodynamic and magnetic properties of nuclei and nuclear matter. In particular, an understanding of how BPS skyrmions respond to an external magnetic field and to pressure would provide us with the corresponding equation of state, which is required for the analysis of nuclear matter in various conditions, from heavy nuclei to neutron stars.
\\
Unfortunately, even the BPS Skyrme model is quite complicated after the minimal $U(1)$ coupling. To overcome the computational difficulties and learn something about the electromagnetic properties of BPS Skyrme type solitons, one can analyze lower-dimensional analogs, as has been done successfully already in many occasions. In fact, there exists a $(2+1)$ dimensional version of the Skyrme model, usually referred to as the baby Skyrme model, which supports solitonic solutions (baby skyrmions) \cite{old}-\cite{shnir} (for the gauged version see \cite{GPS}, \cite{schr1}).  This field theory also possesses its BPS limit, whose Lagrangian consists of the (2+1) dimensional version of the Skyrme term and a potential \cite{GP}-\cite{Sp1}. Moreover, there is again a gauged version of this model, the so-called gauged BPS baby Skyrme model, which has been analyzed recently in the case of an asymptotically vanishing magnetic field \cite{BPS-g}.    

\vspace*{0.2cm}

It is the aim of the present paper to further investigate baby skyrmions in the gauged BPS baby Skyrme model from the perspective of the equation of state for BPS baby skyrmion matter. In particular, we will focus on the issue of how the energy $E$ and volume $V$ of the solitons change if they are put in an asymptotically constant magnetic field $H$ and exposed to external pressure. 

\vspace*{0.2cm}

The paper is organized as follows. In section II we give a general overview on the gauged BPS Skyrme model. We prove the existence of a topological bound for the regularized energy in the case of a non-vanishing but constant asymptotic magnetic field. The BPS equations saturating the bound are presented. In section III we solve the system for the so-called old baby potential, both numerically and analytically in the weak coupling limit. We find the equation of state and related quantities (magnetic compression, magnetization, susceptibility) and prove a ferromagnetic behavior of the BPS baby skyrmion matter. Then, in section IV we introduce pressure and derive the pressure-modified BPS equations. Section V is devoted to the analysis of the equation of state with nonzero pressure and external magnetic field, again for the old baby potential. In section VI we present a toy model for which the equation of state can be obtained analytically for any value of the electromagnetic coupling constant. Finally, we discuss our results. 
%%%%%%%%%%%%%%%%%%%%%%%%%%%%%%%%%%%%%%%%%
\section{The BPS baby Skyrme model in a constant magnetic field}
%%%%%%%%%%%%%%%%%%%%%%%%%%%%%%%%%%%%%%%%%
%%%%%%%%%%%%%%%%%%%%%%%%%%%%%%%%%%%%%%%%%
\subsection{The gauged BPS baby Skyrme model}
%%%%%%%%%%%%%%%%%%%%%%%%%%%%%%%%%%%%%%%%%
Here we briefly summarize the properties of the BPS Skyrme model coupled minimally with the electromagnetic $U(1)$ gauge field.  The model is defined by the following Lagrange density \cite{BPS-g}
\begin{equation}\label{Lag}
 {\cal L} = -\frac{\lambda^2}{4} \left( D_\mu \vec \phi \times D_\nu  \vec \phi \right)^2 - \mu^2 U ( \vec n \cdot \vec \phi)
 + \frac{1}{4 g^2} F^2_{\mu \nu}
\end{equation}
where the covariant derivative reads  \cite{GPS}, \cite{schr1}
\be
D_\mu \vec{\phi}\equiv \partial_\mu \vec{\phi} +A_\mu \vec{n} \times \vec{\phi} .
\ee
Without loss of generality we assume that the constant vector $\vec{n}=(0,0,1)$ and the potential $U$ is a function of the third component of the unit vector field  $\vec{\phi}$. 
The pertinent field equations are
\begin{equation}
 D_\mu \vec{\mathcal{K}}^\mu = -\mu^2 \vec n \times \vec \phi \; U' 
\end{equation}
and the inhomogeneous Maxwell equation is
\begin{equation}
 \partial_\mu F^{\mu \nu} = g^2 \vec n \cdot \vec{\mathcal{K}}^\nu,
\end{equation}
where
\begin{equation}
\vec{\mathcal{K}}^\mu = \lambda^2 D_\nu \vec \phi \left[ \vec \phi \cdot (D^\mu \vec \phi \times D^\nu \vec \phi) \right].
\end{equation}
The full energy functional is
\begin{equation}
E=\frac{1}{2} \int d^2 x \left( \frac{\lambda^2}{2} \left( D_{0} \vec{\phi} \times D_{i} \vec{\phi} \right)^2 + \frac{1}{g^2} E_i^2 + \lambda^2 \left( D_{1} \vec{\phi} \times D_{2} \vec{\phi} \right)^2 +2\mu^2U+\frac{1}{g^2}B^2\right) .
\end{equation}
Further, we assume $\vec n = (0,0,1)$ and the standard axially symmetric static ansatz
\begin{equation} \label{rad-ans}
\vec{\phi} (r,\phi)  = \left( 
\begin{array}{c}
\sin f(r) \cos n\phi \\
\sin f(r) \sin n\phi \\
\cos f(r)
\end{array}
\right), \;\;\;\; A_0=A_r=0, \;\;\; A_\phi=na(r) 
\end{equation}
which leads to an identically vanishing electric field and to the magnetic field $B=\frac{na'(r)}{r}$. Note, that positive $n$ (topological charge) corresponds to a negative magnetic field ($a'$ is always negative as we will see below), while baby anti-skyrmions (negative $n$) would lead to a positive magnetic field. Moreover, we are interested in topologically nontrivial matter field (unit vector field) configurations, which requires the appropriate boundary conditions. $n$ then provides the topological charge (winding number) of $\vec \phi$. The field equations can be rewritten as 
\bea
\frac{1}{r^2} f'' (1+a)^2 \sin^2 f + \frac{f'}{r} \left[ \left( 2a'-\frac{1+a}{r} \right) \frac{1+a}{r} \sin^2 f + \frac{f'}{r} (1+a)^2 \sin f \cos f \right]  && \nonumber \\
+ \; \frac{\mu^2}{n^2\lambda^2} \sin f \; U'&=&0 
\eea
 \begin{equation}
 a''-\frac{a'}{r} = \lambda^2 g^2 (1+a) \sin^2 f f'^2 
\end{equation} 
where now $U=U(\phi_3) = U(\cos f)$ and $U' = U_{\phi_3}$. It is also convenient to introduce the new variable 
\begin{equation}
y=\frac{r^2}{2}
\end{equation}
which allows to rewrite the equations as  the following system of autonomous second order equations 
\begin{equation}
\sin f \left\{ \partial_y \left[ f_y (1+a)^2\sin f\right] + \frac{\mu^2}{n^2 \lambda^2} U' \right\}=0
\end{equation} 
\begin{equation}
a_{yy}=\lambda^2g^2(1+a)\sin^2 f f_y^2 .
\end{equation} 
Further, introducing a new target space variable $h$ 
\begin{equation}
\phi_3 = \cos f \equiv 1-2h \; \Rightarrow \; h =\frac{1}{2} (1-\cos f), \;\;\; h_y = \frac{1}{2} \sin f f_y
\end{equation} 
this may be further simplified to
\begin{equation} \label{h-eq}
\sin f \left\{ \partial_y \left[ h_y (1+a)^2 \right] - \frac{\mu^2}{4n^2 \lambda^2} U_h \right\}=0
\end{equation} 
\begin{equation} \label{a-eq}
a_{yy}=\lambda^2g^2(1+a) 4 h_y^2 
\end{equation} 
where now $U=U(h)$ and $U_h =-2U'$. It has been previously found that the model preserves many properties of the original ungauged version \cite{GP}, \cite{restr-bS}, \cite{Sp1}. 
\\
First of all, there is a BPS bound which can be saturated by the corresponding BPS configurations. The important assumption in the proof was the boundary condition for the magnetic field that it asymptotically vanishes. Then, the energy is bounded from the below by
\be
E \geq 4\pi E_0 \lambda^2 |k|  <W'>_{S^2}
\ee 
where the inequality is saturated for the pertinent BPS solutions. Here $k$ is the topological charge (winding number) and  $<W'>_{S^2}$ is the average value of the derivative of the superpotential (see below) over the target space manifold. The resulting BPS baby skyrmions may be of the compacton type with the magnetic field completely confined inside the compact baby skyrmions. Further, the flux is not quantized (except in the large $g$ limit). One interesting conjecture, verified in many particular examples, was the absence of gauged BPS baby skyrmions for potentials with more than one vacuum. This strongly differs from the ungauged case where such topological solitons do exist. 
\\
Secondly, the model is integrable in the sense of generalized integrability \cite{gen-int} (no conditions for the gauge field) which means that there are infinitely many conservation laws (genuine conservation laws, which are not related to the gauge transformations). Moreover, the static energy functional possesses the area preserving diffeomorphisms as its symmetry group. Therefore, the moduli space of BPS solutions is infinite-dimensional. This also means that our assumed ansatz does not restrict the form of the solutions. One may use the base space area preserving diffeomorphisms to construct solutions with arbitrary (not axially symmetrical) shapes.  

%%%%%%%%%%%%%%%%%%%%%%%%%%%%%%%%%%%%%%%%%
\subsection{Constant asymptotical magnetic field}
%%%%%%%%%%%%%%%%%%%%%%%%%%%%%%%%%%%%%%%%%
The problem we want to solve next is how the external constant magnetic field $H$ modifies the BPS gauged baby skyrmions originally obtained in 
\cite{BPS-g}. Obviously, the field equations remain unchanged 
\begin{equation} \label{hh-eq}
\partial_y \left[ h_y (1+a)^2 \right] - \frac{\mu^2}{4n^2 \lambda^2} U_h =0
\end{equation} 
\begin{equation} \label{aa-eq}
a_{yy}=\lambda^2g^2(1+a) 4 h_y^2 ,
\end{equation} 
but the boundary conditions are different. Now,
\be
h(y=0)=1, \;\;\; a(y=0)=0
\ee
\be
h(y=y_0)= h_y(y=y_0)=0, \;\;\; a_y(y=y_0)=\frac{H}{n}
\ee
where the last condition leads to an asymptotically constant magnetic field $B(y=y_0)=H ={\rm const.}$. Here, $y_0$ can be finite (compactons - for example in the case of the old baby Skyrme potential) or infinite. As the zero boundary conditions played a crucial role for the proof of the existence of the BPS bound, as well as for its saturation by solutions of the BPS equations, it is not obvious whether all these properties survive after the change of the boundary conditions.  
Here we restrict ourselves to $n>0$. The corresponding analysis for negative topological charge is straightforward and requires the interchange of $H$ to $-H$. 
%%%%%%%%%%%%%%%%%%%%%%%%%%%%%%%%%%%%%%%%%
\subsection{The BPS bound for constant asymptotical magnetic field}
%%%%%%%%%%%%%%%%%%%%%%%%%%%%%%%%%%%%%%%%%
Here we would like to derive a BPS bound in the case of an asymptotically constant magnetic field. This requires some important improvements in the original derivation. 
Consider the following non-negative integral  
\begin{equation}
0 \leq \frac{1}{2} E_0 \int d^2 x \left[ \lambda^2 (Q-w(\phi_3))^2 + \frac{1}{g^2} (B+b(\phi_3))^2\right] = 
\end{equation}
\begin{equation}
= \frac{1}{2} E_0 \int d^2 x \left[ \lambda^2 Q^2 + \lambda^2 w^2+\frac{1}{g^2} B^2+\frac{1}{g^2} b^2 -2\lambda^2 qw - 2\lambda^2\epsilon_{ij} A_i\partial_j \phi_3 w +\frac{2}{g^2} \epsilon_{ij}\partial_i A_j b \right] \label{bound1}
\end{equation}
where $b$ and $w$ are (at the moment arbitrary) functions of the field variable $\phi_3$. Further, 
\begin{equation}
Q = q+\epsilon_{ij} A_i \partial_j (\vec{n} \cdot \vec{\phi} ), \;\;\; q = \vec{\phi} \cdot \partial_1 \vec{\phi} \times \partial_2 \vec{\phi} .
\end{equation}
Now, let
\begin{equation}
b(\phi_3)=g^2\lambda^2 W - H, \;\;\; W \equiv \int_{\phi_{3,v}}^{\phi_3} dt w(t)
\end{equation}
where $H$ is a constant equal to the asymptotic value of the magnetic field. Further, the "superpotential" $W$ is a function of the field variable which depends on the potential $U$ (see Eq. (\ref{superpot-eq})), as we shall see in a moment. 
\\
The last terms in (\ref{bound1}) can be written as
\begin{equation}
E_0\int d^2x \left[ \lambda^2 \epsilon_{ij} \partial_i (A_j W) - \frac{H}{g^2} \epsilon_{ij} \partial_i A_j \right]  = -E_0\int d^2x \frac{1}{g^2} BH
\end{equation}
as the first part vanishes at the compacton boundary where $W(\phi_{3,v}) =0$ by definition. Then
\begin{equation}
0 \leq  \frac{1}{2} E_0 \int d^2 x \left[ \lambda^2 Q^2 + \frac{1}{g^2} B^2 +\lambda^2 W'^2 + g^2\lambda^4 W^2 -2\lambda^2WH \right] - E_0 \lambda^2 \int d^2 x qW' 
\end{equation}
\begin{equation}
+\frac{1}{2} E_0 \int d^2 x \frac{1}{g^2} (H^2-2HB) .
\end{equation}
Hence,
\begin{equation}
\frac{1}{2} E_0 \int d^2 x \left[ \lambda^2 Q^2 + \frac{1}{g^2} B^2 +2\mu^2 U \right] \geq  E_0 \lambda^2 \int d^2 x qW' -\frac{1}{2} E_0 \int d^2 x \frac{1}{g^2} (H^2-2HB)
\end{equation}
i.e.,
\begin{equation}
\frac{1}{2} E_0 \int d^2 x \left[ \lambda^2 Q^2 + \frac{1}{g^2} (B-H)^2+2\mu^2 U \right] \geq  E_0 \lambda^2 \int d^2 x qW' 
\end{equation}
where the superpotential equation relating the potential $U$ and the superpotential $W$ reads
\begin{equation} \label{superpot-eq}
\lambda^2 W'^2 + g^2\lambda^4 W^2 -2\lambda^2WH =2\mu^2 U ,
\end{equation}
which differs from the expression found in \cite{BPS-g} for zero asymptotic magnetic field by the term linear in $W$ (and in $H$). 
By construction, $W(\phi_3=1)=0$, which leads to $W'(\phi_3=1)=0$.
Let us remark that this new superpotential equation can be brought to the form of the original superpotential equation by the following redefinition
\be
\tilde W=W-\frac{1}{g^2\lambda^2}H, \;\;\; \tilde U = U+\frac{1}{2g^2\mu^2}H^2 .
\ee
Then 
\be
\lambda^2 \tilde{W}'^2 + g^2\lambda^4 \tilde{W}^2 =2\mu^2 \tilde U.
\ee
However, now the boundary conditions for the superpotential $\tilde W$ are changed.
\\
It is convenient to define a regularized energy where we subtract the infinite contribution from the asymptotically constant magnetic field
\be
E_{reg}= \frac{E_0}{2} \int d^2 x \left[ \lambda^2 \left( D_1\vec{\phi} \times D_2 \vec{\phi}\right)^2 +2\mu^2 U +\frac{1}{g^2} (B-H)^2
\right] .
\ee
Then
\begin{equation} \label{BPSbound}
E_{reg} \geq E_0 \lambda^2 \int d^2 x qW' \equiv 4\pi |k| E_0 \lambda^2 <W'>_{S^2} .
\end{equation}
Obviously, the inequality is saturated if 
\begin{equation}
Q= W' \label{bps1}
\end{equation}
\begin{equation}
B = - g^2\lambda^2W +H \label{bps2}
\end{equation}
which are the BPS equations in the case of a constant asymptotic magnetic field. For the shifted superpotential we get the usual form of the BPS equations
\begin{equation}
Q= \tilde W' \label{bps1-t}
\end{equation}
\begin{equation}
B = - g^2\lambda^2\tilde W . \label{bps2-t}
\end{equation}
It remains to be shown that the solutions of these equations obey the full second order equations of motion,
\begin{equation}
\lambda^2 \epsilon_{ij} D_i [(D_j \vec{\phi}) Q] = -\mu^2 U' \vec{n} \times \vec{\phi}
\end{equation}
\begin{equation}
\partial_i F^{ij} = g^2\lambda^2 \vec{n} \cdot D^k \vec{\phi} (\vec{\phi} \cdot D^j \vec{\phi} \times D_k \vec{\phi} ) .
\end{equation}
The Maxwell equation follows in the same way as in the $H=0$ case since the derivative of (\ref{bps2}) does not depend on the value of $H$. 
\\
Further, from the superpotential equation we get
\begin{equation}
\mu^2 U'=\lambda^2 W'W''+g^2\lambda^4 WW' -\lambda HW'
\end{equation}
and
\begin{equation}
\partial_k Q = W'' \partial_k (\vec{n} \cdot \vec{\phi}) .
\end{equation}
And then we follow the same derivation as in the $H=0$ case. Namely, rewriting the first equation of motion as 
\begin{equation}
D_2 \vec{\phi} \partial_1 Q - D_1 \vec{\phi} \partial_2 Q +\vec{n} \times \vec{\phi} BQ=-\lambda^{-2} \mu^2 U' \vec{n} \times \phi
\end{equation}
and using the above formulas we get
\begin{equation}
(D_2 \vec{\phi} \partial_1 (\vec{n} \cdot \vec{\phi}) - D_1 \vec{\phi}\partial_2 (\vec{n} \cdot \vec{\phi})) W'' = \vec{n} \times \vec{\phi} \left( g^2\lambda^4 WW' -\lambda HW' - W'W'' -g^2\lambda^4WW' +\lambda W'H\right)
\end{equation}
i.e., 
\begin{equation}
D_2 \vec{\phi} \partial_1 (\vec{n} \cdot \vec{\phi}) - D_1 \vec{\phi} \partial_2 (\vec{n} \cdot \vec{\phi})  = - \vec{n} \times \vec{\phi} W'
\end{equation}
which is the same as for $H=0$. The remaining steps: using the covariant derivative definition, use  $Q=W'$ and the definition of $Q$, do not depend on $H$. That ends the proof.  
\\
Finally, let us observe that in the axially symmetric ansatz the BPS equations read
\begin{equation} \label{ax-BPS1}
2nh_y (1+a) = - \frac{1}{2} W_h
\end{equation}
\begin{equation}
na_y=-g^2\lambda^2 W + H
\end{equation}
or for the shifted superpotential
\begin{equation}
2nh_y (1+a) = - \frac{1}{2} \tilde W_h
\end{equation}
\begin{equation}
na_y=-g^2\lambda^2 \tilde W .
\end{equation}
%%%%%%%%%%%%%%%%%%%%%%%%%%%%%%%%%%%%%%%%%
\subsection{The regularized flux}
%%%%%%%%%%%%%%%%%%%%%%%%%%%%%%%%%%%%%%%%%
Another important quantity is the flux of the magnetic field
\be
\Phi = \int rdr d\phi B .
\ee 
As the magnetic field extends to infinity the flux will also take an infinite value. However, for compactons, which is the case discussed in the paper, the magnetic field outside the solitons is exactly equal to the external field. Due to that we are rather interested in the value of the flux integrated over the area of the solitons, which is equivalent (up to an additive constant) to the following definition of 
the regularized flux 
\be
\Phi_{reg}=\int rdr d\phi (B-H) = 2\pi \int_0^{r_0} rdr (B-H) 
\ee
where the axially symmetric configuration has been assumed.
Then, using the definition of the magnetic field and the behavior at the boundary we find
\be
\Phi_{reg}=2\pi n \int dy \left(a_y -\frac{H}{n} \right) = 2\pi n \int dy \partial_y \left(a -\frac{Hy}{n} \right) = 2\pi n  \left(a(y_0) -\frac{Hy_0}{n} \right) .
\ee
It is also possible to prove that this value depends only on the model (coupling constants and the form of the potential) but not on the local behavior of a particular solution. Dividing one BPS equation by the other we find
\be
\frac{a_y}{1+a}=\frac{4(g^2\lambda^2W -H)h_y}{W_h}
\ee
i.e.,
\be
\partial_y \ln (1+a) = \partial_y F
\ee
where
\be
 F_h=\frac{4(g^2\lambda^2 W-H)}{W_h} \;\; \Rightarrow \;\; F(h)=\int_0^{h} dh'\frac{4(g^2\lambda^2W(h') -H)}{W_{h'}(h')} .
\ee
Then, 
\be
\ln C (1+a) = F(h(y))
\ee
where the constant $C$ can be computed from the boundary values of the fields at $y=0$, 
\be
C = e^{F(h=1)}.
\ee
Therefore, we get
\be \label{sol-a(y)}
a(y)= e^{F(y)-F(1)} -1
\ee
and, specifically at $y=y_0$ where, by definition, $F(h=0) \equiv 0$,
\be
a (y_0) = -1+e^{-F(1)} = -1+e^{-g^2\lambda^2 A+HB}
\ee
where the constants $A,B$ depend on the model (potential),
\be
A=\int_0^{1} dh \frac{4W(h) }{W_{h}(h)}, \;\;\; B=\int_0^{1}\frac{dh}{W_{h}(h)} .
\ee
It is clear that $a(y_0) \rightarrow -1$ once $g \rightarrow \infty$ or $H \rightarrow -\infty$. This behavior is confirmed by numerical results.

For the regularized flux we then get
\be
\Phi_{reg}= 2\pi n (-1+e^{-F(1)}) -HV
\ee
where $V=2\pi y_0$ is the "volume" (area) of the compacton. We use the word "volume" and the letter $V$ to maintain close contact with the standard thermodynamic notation. We already showed that the first part, $2\pi n(-1+\exp (-F(1))$, may be expressed as a target space integral and, therefore, does not depend on the specific solution $h(y), a(y)$. In other words, it is one and the same thermodynamic function for all equilibrium configurations (BPS solutions). In a next step, let us demonstrate that also the "volume" $V$ (and, consequently, the full regularized flux) is a thermodynamic function, i.e., a given function of $H$ for all BPS solutions.  
The BPS equation (\ref{ax-BPS1}) may be re-expressed like
\be
dy = -4n \frac{1+a}{W_h} dh = - 4n\frac{e^{F(h)-F(1)}}{W_h} dh
\ee
where we used (\ref{sol-a(y)}) in the second step. Integrating both sides over their respective ranges leads to
\be \label{vol}
V(H) = 2\pi y_0 = 8\pi e^{-F(1)} \int _0^1 dh \frac{e^{F(h)}}{W_h}
\ee
and to the regularized flux
\be
\Phi_{reg}= 2\pi n \left( -1+e^{-F(1)} - 4He^{-F(1)}  \int _0^1 dh \frac{e^{F(h)}}{W_h}  \right) 
\ee
which, indeed, is a thermodynamic function, as announced.

%%%%%%%%%%%%%%%%%%%%%%%%%%%%%%%%%%%%%%%%%
\subsection{The magnetization}
%%%%%%%%%%%%%%%%%%%%%%%%%%%%%%%%%%%%%%%%%
The thermodynamic magnetization $M$ is defined as minus the change of the thermodynamic energy of a sample (in our case, the skyrmion) under a variation of the external magnetic field. Here, the electromagnetic part of the thermodynamic energy must be calculated from the difference of the electromagnetic fields with and without the sample, which precisely corresponds to our definition of the regularized energy, i.e.,
\be
M=-\frac{\partial E_{\rm reg}}{\partial H} .
\ee    
We use the BPS bound (\ref{BPSbound}) for the energy and express the average value of $W'$ over the target space $\mathbb{S}^2$ like
\be
\langle W' \rangle \equiv \langle W_{\phi_3}\rangle = \frac{1}{4\pi} \int_0^{2\pi} d\varphi \int df \sin f W_{\phi_3} = \frac{1}{2} \int_{-1}^1 d\phi_3  
W_{\phi_3}=\frac{1}{2}\int_0^1 dh W_h = \frac{1}{2}W(h=1)
\ee
where we treated $W$ as a function of $h$ in the last two terms, which we shall continue to do, i.e., $W(1)\equiv W(h=1)$ in what follows. The magnetization then is
\be
M= -2\pi n \lambda^2 \frac{ \partial W(1)}{\partial H}
\ee
and, obviously, is a thermodynamic function (i.e., the same function of $H$ for all equilibrium configurations). 

In standard thermodynamics there is a simple relation between the magnetization and the difference between full and external magnetic flux in the sample. In our conventions, this relation reads
\be \label{M-phi_reg}
M = \frac{1}{g^2}\int (B-H) \equiv \frac{1}{g^2}\Phi_{\rm reg}.
\ee
We shall see that this relation continues to hold in our model, although the proof is not trivial and makes use of the BPS nature of the model, specifically of the superpotential equation. Using the variable $h$ instead of $\phi_3$, the superpotential equation may be re-expressed like
\be
\frac{1}{4}W_h^2 + \tilde g^2 W^2 - 2WH = 2\tilde \mu^2 U(h) \; , \quad \tilde g = \lambda g \; ,\quad \tilde \mu = \frac{\mu}{\lambda}.
\ee
To express the first derivative $\partial_H W(1)$, it is useful to introduce a first order (infinitesimal) shift about a given value $H_0$,
\be
H=H_0 + \delta \; ,\quad W = W^{(0)} +  W^{(1)} \delta + {\rm \bf O}(\delta^2 )
\ee
 then the magnetization at $H=H_0$ is
\be
M(H_0)= -2\pi n \lambda^2 W^{(1)}(1)
\ee
and the thermodynamic relation (\ref{M-phi_reg}) becomes
\be \label{M-phi_reg-1}
-2\pi n \lambda^2 W^{(1)}(1) = \Phi_{\rm reg} (H_0) .
\ee
The superpotential equation at zeroth order in $\delta$ is
\be
\frac{1}{4}(W^{(0)}_h)^2 + \tilde g^2 (W^{(0)})^2 - 2W^{(0)}H_0 = 2\tilde \mu^2 U(h) 
\ee
and serves to determine $W^{(0)}$ for a given $H_0$, potential $U$ and given coupling constants. The first order superpotential equation is (remember that $U$ does not depend on $H$)
\be
\frac{1}{2} W_h^{(0)}W_h^{(1)} + 2 \tilde g^2 W^{(0)} W^{(1)} - 2 H_0 W^{(1)} - 2 W^{(0)} =0
\ee
 or
\be
\frac{1}{4}\frac{W_h^{(0)}}{\tilde g^2 W^{(0)} - H_0} W_h^{(1)} + W^{(1)} = \frac{W^{(0)}}{\tilde g^2 W^{(0)} - H_0}
\ee
and serves to determine $W^{(1)}(h)$ for a given $W^{(0)}(h)$. Indeed, introducing a new variable
\be
k = F(h) = 4\int_0^h dh' \frac{\tilde g^2 W^{(0)}(h') - H_0}{W_{h'}^{(0)}}
\ee
the above equation becomes
\be
W_k^{(1)} + W^{(1)} = \frac{W^{(0)}}{\tilde g^2 W^{(0)} - H_0}
\ee
and may be easily solved via the method of the variation of the integration constant, leading to
\be
W^{(1)}(k) = c(k) e^{-k} \; , \quad c(k) = \int_0^k dk' e^{k'}\frac{W^{(0)}}{\tilde g^2 W^{(0)} - H_0}
\ee
or, in terms of the variable $h$
\be
W^{(1)}(h) = 4 e^{-F(h)} \int_0^h dh' \frac{W^{(0)}}{W^{(0)}_{h'}}e^{F(h')} .
\ee
In particular, for $W^{(1)}(1)$ we find 
\bea
W^{(1)} (1) &=& 4e^{-F(1)} \int_0^1 dh \frac{W^{(0)}}{W^{(0)}_h} e^{F(h)} \nonumber \\
&=& \frac{e^{-F(1)}}{\tilde g^2} \left( \int_0^1 dh \frac{4(\tilde g^2 W^{(0)} -H_0)}{W^{(0)}_h}e^{F(h)} + 4H_0 \int_0^1 dh \frac{e^{F(h)}}{W^{(0)}_h} \right) \nonumber \\
&=& \frac{e^{-F(1)}}{\tilde g^2} \left( -1+e^{F(1)} + 4H_0 \int_0^1 dh \frac{e^{F(h)}}{W^{(0)}_h} \right) 
\eea
where 
\be
de^{F(h)} = \frac{4(\tilde g^2 W^{(0)} -H_0)}{W^{(0)}_h}e^{F(h)} dh
\ee
and $F(0)=0$ was used. From this last result, the thermodynamic relation (\ref{M-phi_reg-1}) follows immediately.

%%%%%%%%%%%%%%%%%%%%%%%%%%%%%%%%%%%%%%%%%
\section{Constant magnetic field and the old baby potential}
%%%%%%%%%%%%%%%%%%%%%%%%%%%%%%%%%%%%%%%%%
%%%%%%%%%%%%%%%%%%%%%%%%%%%%%%%%%%%%%%%%%
\subsection{Numerical computations}
%%%%%%%%%%%%%%%%%%%%%%%%%%%%%%%%%%%%%%%%%
\begin{figure}
\includegraphics[height=4.9cm]{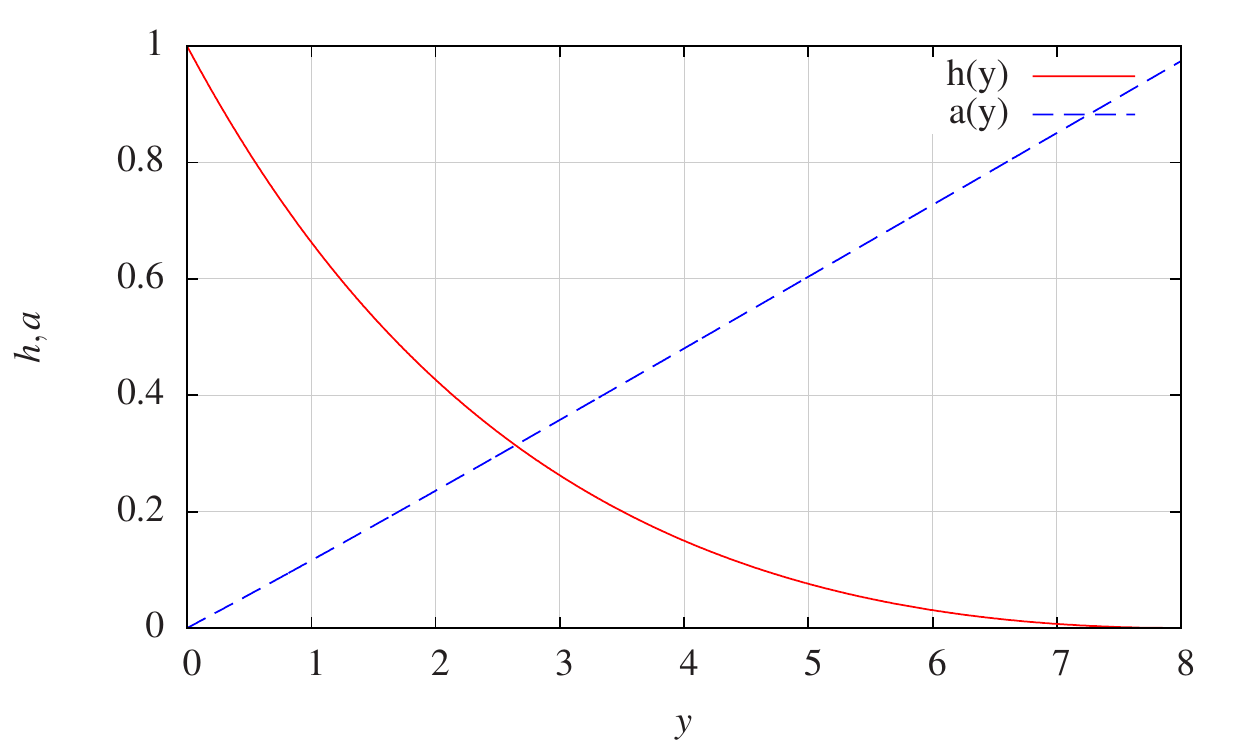}
\includegraphics[height=4.9cm]{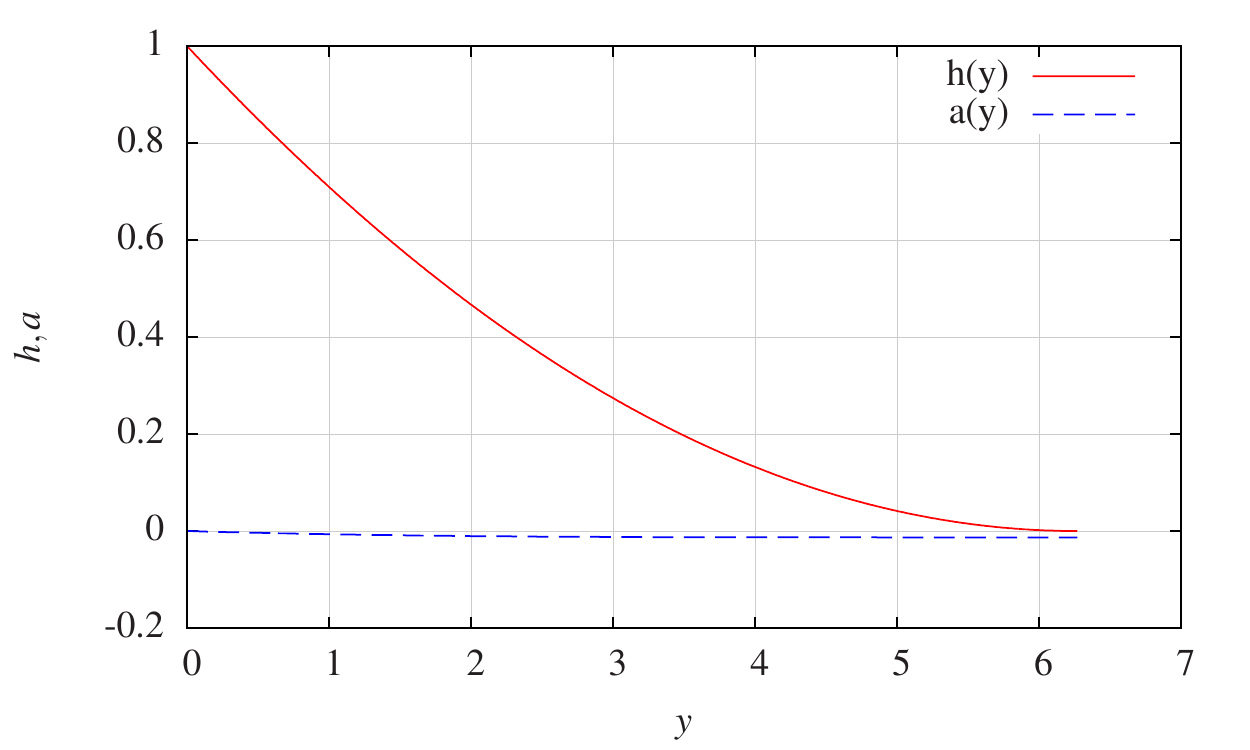}
\includegraphics[height=4.9cm]{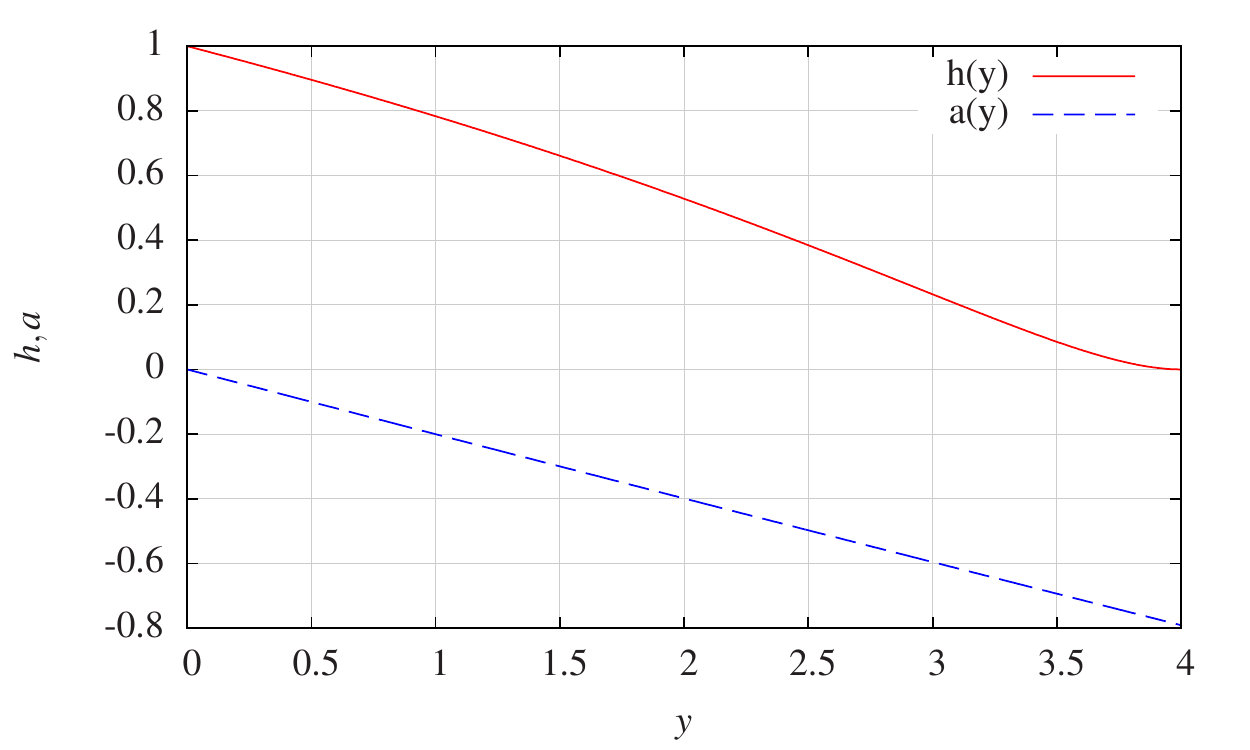}
\includegraphics[height=4.9cm]{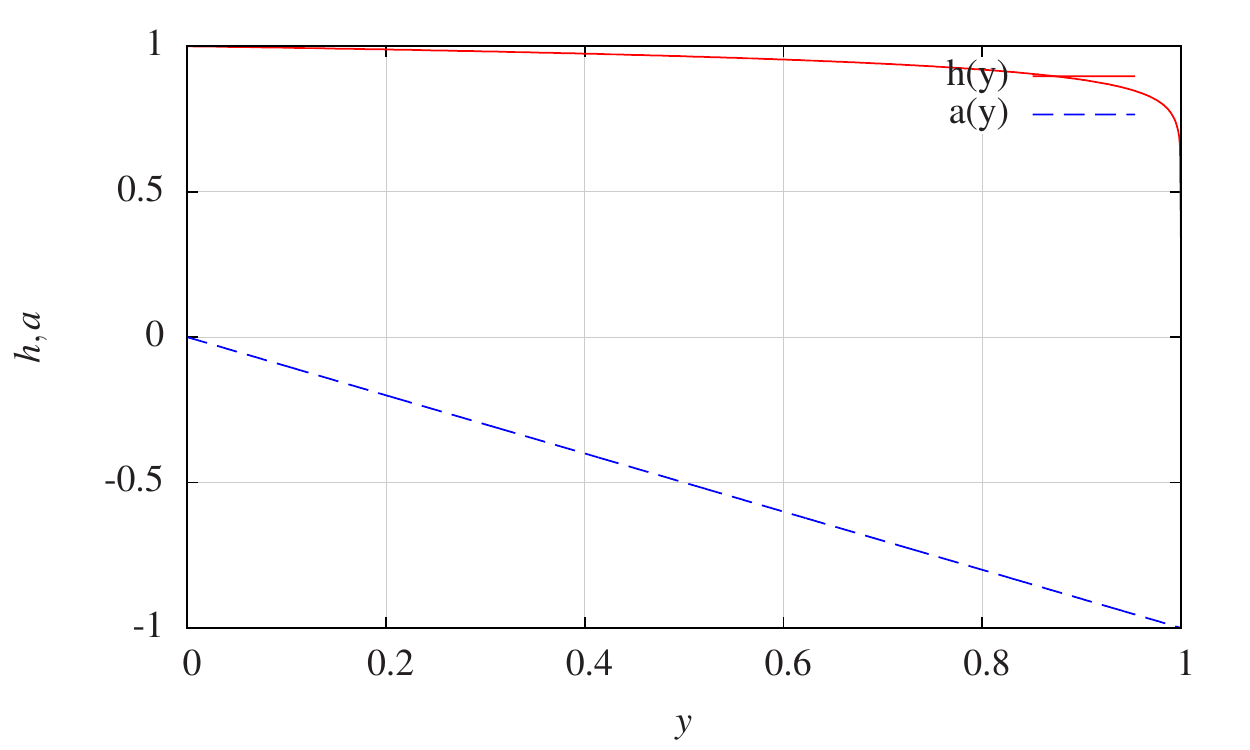}
\caption{The baby skyrmion profile $h$ and the magnetic potential $a$ for $H=0.1236, -1.167 \cdot 10^{-6}, -0.1952, -0.9987$ and $g=0.1$}
\end{figure}
\begin{figure}
\includegraphics[height=5.9cm]{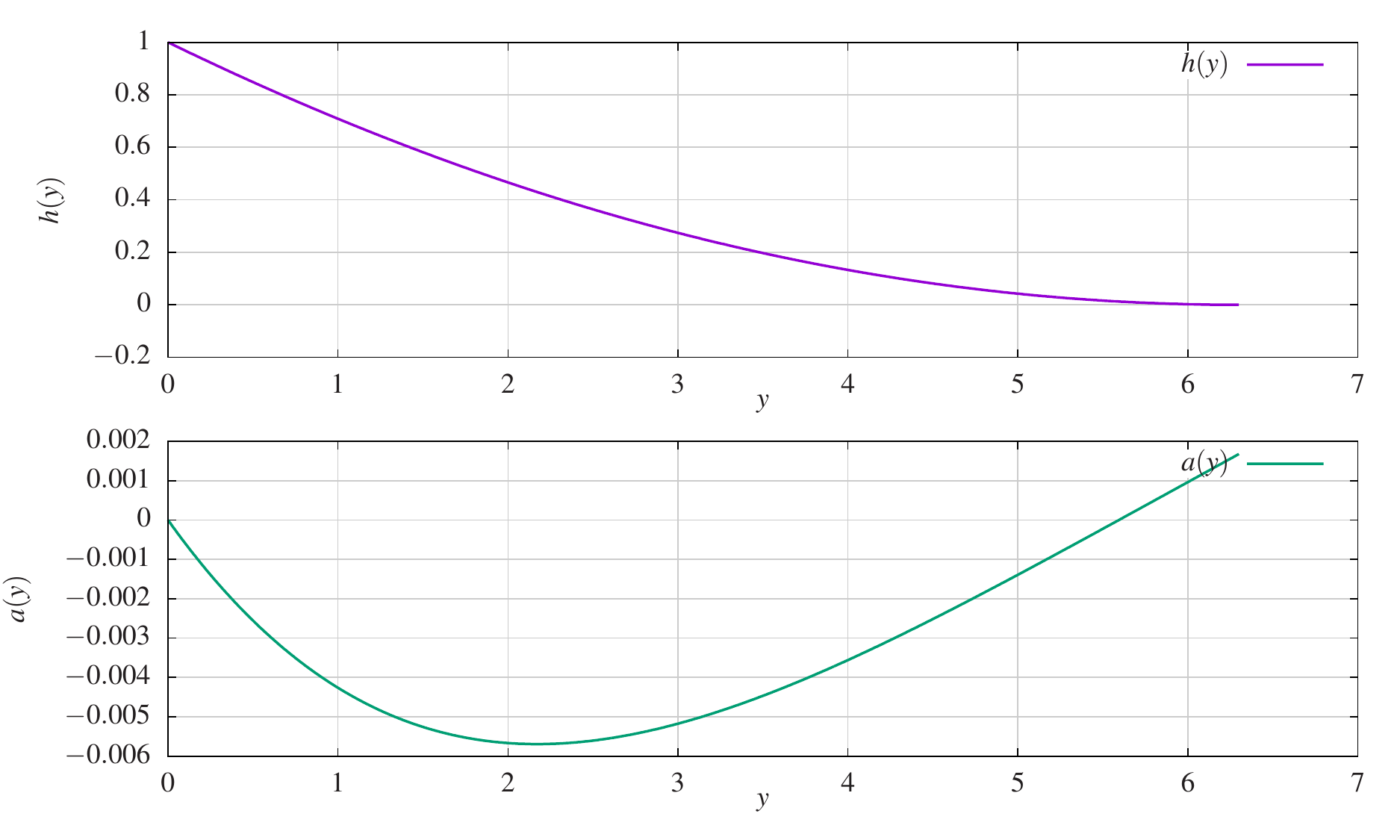}
\caption{The baby skyrmion profile $h$ and the magnetic potential $a$ for $H=0.002378 $ and $g=0.1$.  }
\end{figure}
The system introduced above significantly simplifies in the case of the old baby Skyrme potential
\be \label{old-pot}
U=1-\phi^3 \;\;\;\; \Rightarrow \;\;\;\; U(h)=2h.
\ee
Then the field equations can be integrated to 
\be
h_y (1+a)^2 = \frac{\mu^2}{2n^2\lambda^2} (y-y_0)
\ee
and 
\be
(1+a)^3a_{yy}=\frac{g^2\mu^4}{n^4\lambda^2} (y-y_0)^2 .
\ee
The corresponding energy integral is
\be
E= 2\pi \int dy \left( 2\lambda^2n^2(1+a)^2h_y^2 +2\mu^2h +\frac{1}{2g^2} n^2a_y^2\right) .
\ee
Effectively, the problem depends on two coupling constants. The dependence on the topological charge can be included into a redefinition of the base space coordinate while a particular value of $\lambda$ just fixes the energy scale. So, let us choose $n=1, \lambda=1$ and treat $\mu$ and $g$ as parameters (now dimensionless) defining different  theories. Moreover, the external magnetic field $H$ is another free parameter. 
\begin{figure}
%\hfill
\includegraphics[height=6.6cm]{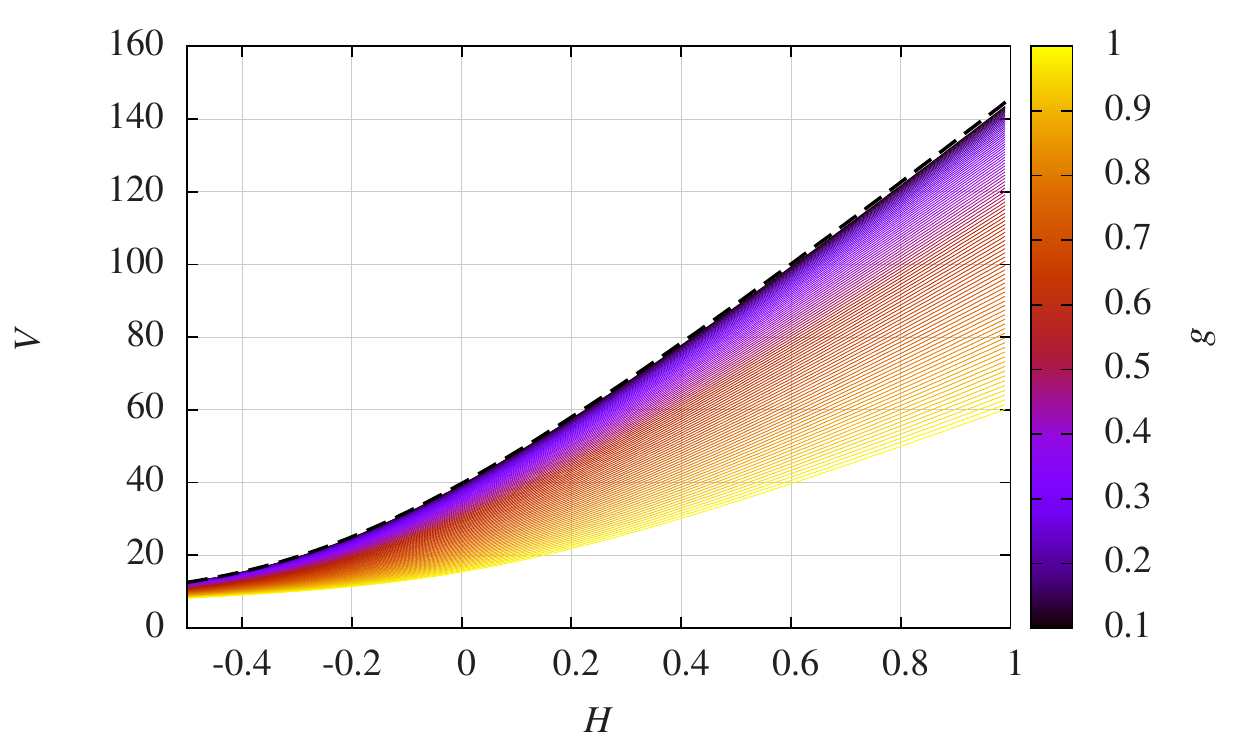}
% \hfill
\caption{Dependence of the compacton "volume" (more precisely: area) on the constant asymptotic magnetic field $H$ for different values of the coupling constant $g$. The non-back reaction  approximation is denoted by a dashed line.}
\end{figure}
\begin{figure}
%\hfill
\includegraphics[height=4.9cm]{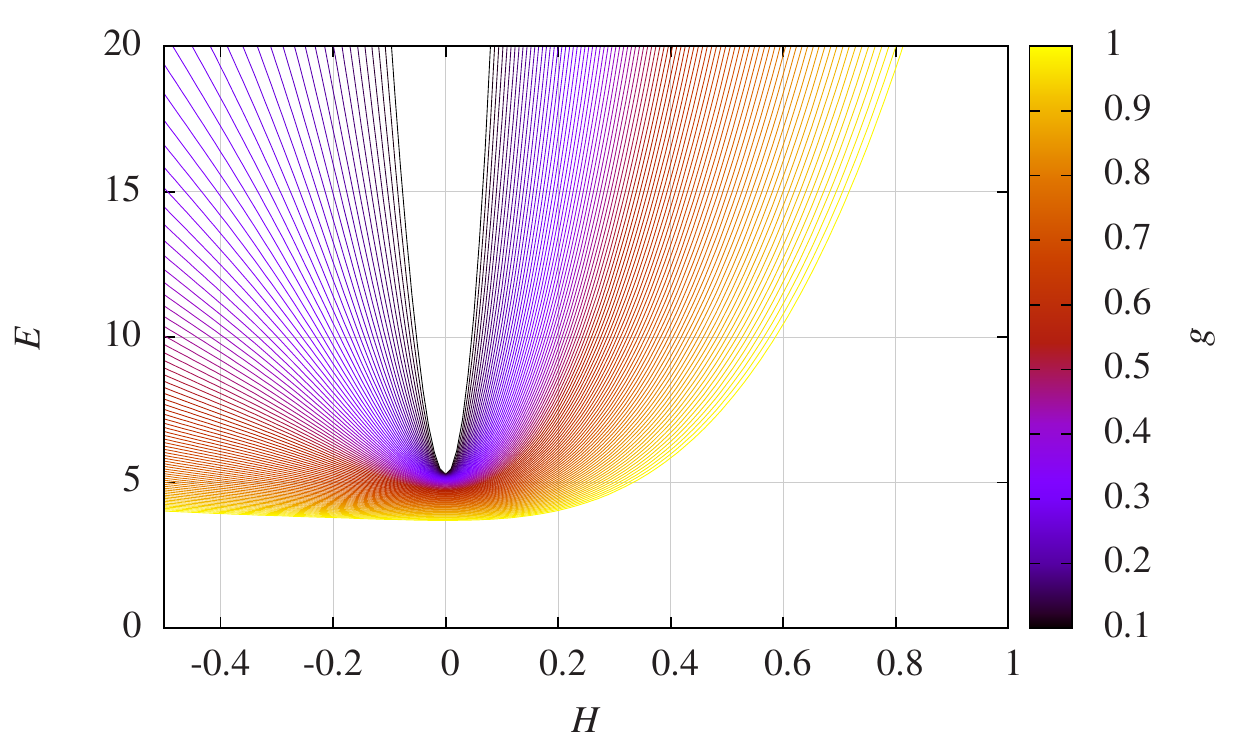}
\includegraphics[height=4.9cm]{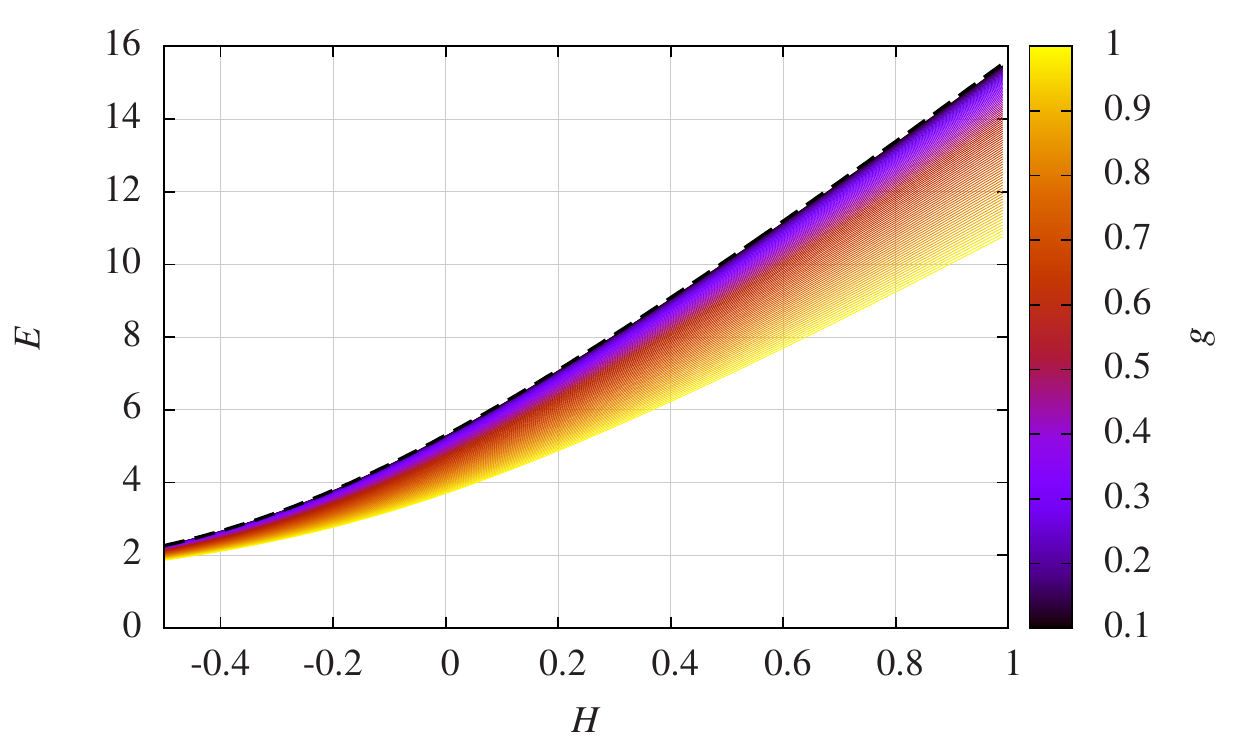}
% \hfill
\caption{Dependence of  the total energy inside the compacton domain (left figure) and of the regularized energy (right figure) on the constant asymptotic magnetic field $H$, for different values of the coupling constant $g$. The non-back reaction  approximation is denoted by a dashed line.}
\end{figure}
\\
As in the $H=0$ case we expand the functions at the boundary 
\be
h=\frac{\mu^2}{4n^2\lambda^2(1+b_0)^2} (y-y_0)^2+....
\ee
\be
a= b_0+\frac{H}{n} (y-y_0) + \frac{g^2\mu^4}{12n^4\lambda^2(1+b_0)^3} (y-y_0)^4+...
\ee
In the numerical computations we assumed $\mu^2=0.1$ (the results for $\mu^2=1$ and $\mu^2=10$ are very similar) and then looked for a few different $g$ and scanned for a wide range of $H$. 

Examples of gauged BPS baby skyrmions are plotted in Fig. 1 for different values of the external magnetic field. The electromagnetic coupling constant is $g=0.1$.  
At this point it is useful to remember that the gauged baby BPS skyrmions without external magnetic field have a magnetic field which is everywhere negative (for positive bayon number $n$) and a negative magnetization proportional to the baryon number \cite{BPS-g}. In other words, these gauged skyrmions show a ferromagnetic behaviour. For a negative external field we therefore expect that the negative magnetic field will become stronger (i.e., more negative). As the gauge potential for negative magnetic field is restricted to the interval $a(y) \in (-1,0]$, as follows easily from eq. (\ref{sol-a(y)}), the stronger (more negative) magnetic field is achieved by shrinking the size of the skyrmion. Concretely, for strong  
negative $H\ll 0$ we approach a singular configuration: the skyrmion profile gets flatter and flatter inside (approximately constant charge density) with a rapid but smooth approach to the vacuum at the boundary whereas $a$ has a more and more linear dependence on $y$ tending to $a_\infty=-1$. In the limit where $H \rightarrow - \infty$ the size of the compacton goes to 0 as $y_0 \sim \frac{1}{|H|}$ and the solutions approach the step function and a linear function for $h$ and $a$, respectively. The approach to the limiting step function solution is faster for higher values of the electromagnetic coupling constant $g$. 

For high positive values of $H$, the magnetic field changes sign everywhere, and the resulting  gauge potential $a$ is a simple monotonously increasing function from 0 to $a_\infty>0$.  For a positive but sufficiently small $H$, however, the phenomenon of magnetic flux inversion occurs. That is to say, the magnetic field $B(y)$ is negative in a ball $0\le y <y_*$ (because the magnetic field without exterenal field is more negative in the core region), becomes zero at $y_*$ and positive in the shell $y_* <y\le y_0$ (because $B(y_0)=H$ must hold at the compacton boundary).   The corresponding gauge potential is, therefore, a decreasing function in the ball close to the center  but an increasing function in the shell.
Finally, the value of the gauge potential at the compacton boundary $a(y_0)$ determines the total magnetic flux inside the compacton.
Specifically, the total magnetic flux inside the compacton may become zero, in constrast to the regularized flux or magnetization, which is always  negative for positive baryon number.
The baby skyrmion profile $H$ is a simple monotously decreasing function for all values of $H$.  We show an example of the magnetic flux inversion in Fig. 2.

In Fig. 3 and Fig. 4 we show how the compacton size and the compacton energy, respectively, depend on the external magnetic field.   

\begin{figure}
\includegraphics[height=4.9cm]{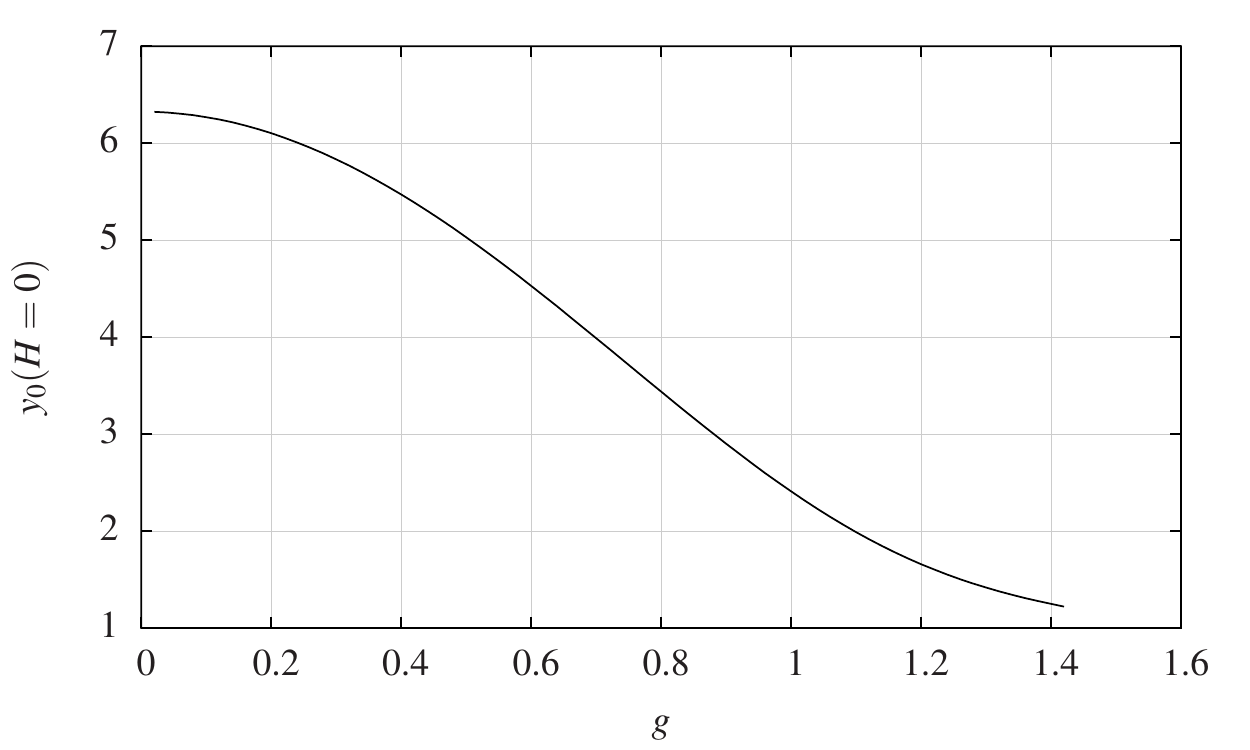}
\includegraphics[height=4.9cm]{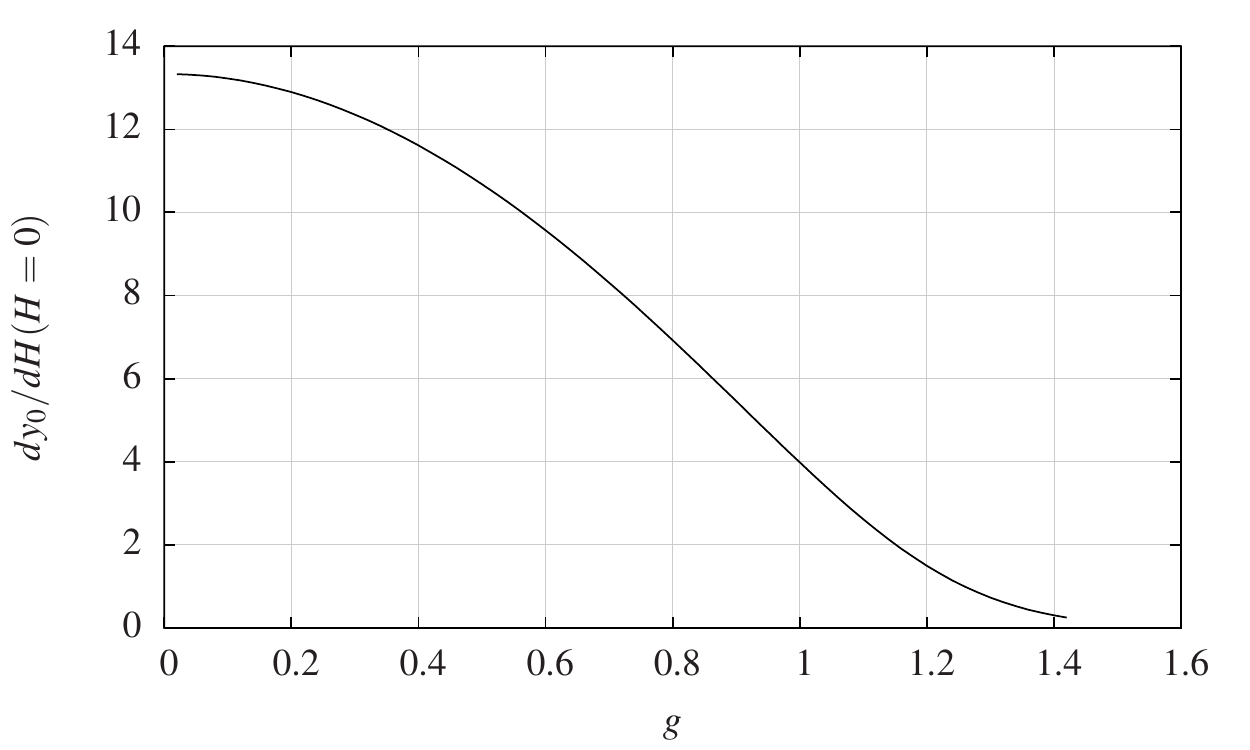}
\caption{The size of the skyrmions and its derivative as a function of $g$ at $H=0$.}
\end{figure}
\begin{figure}
\includegraphics[height=4.9cm]{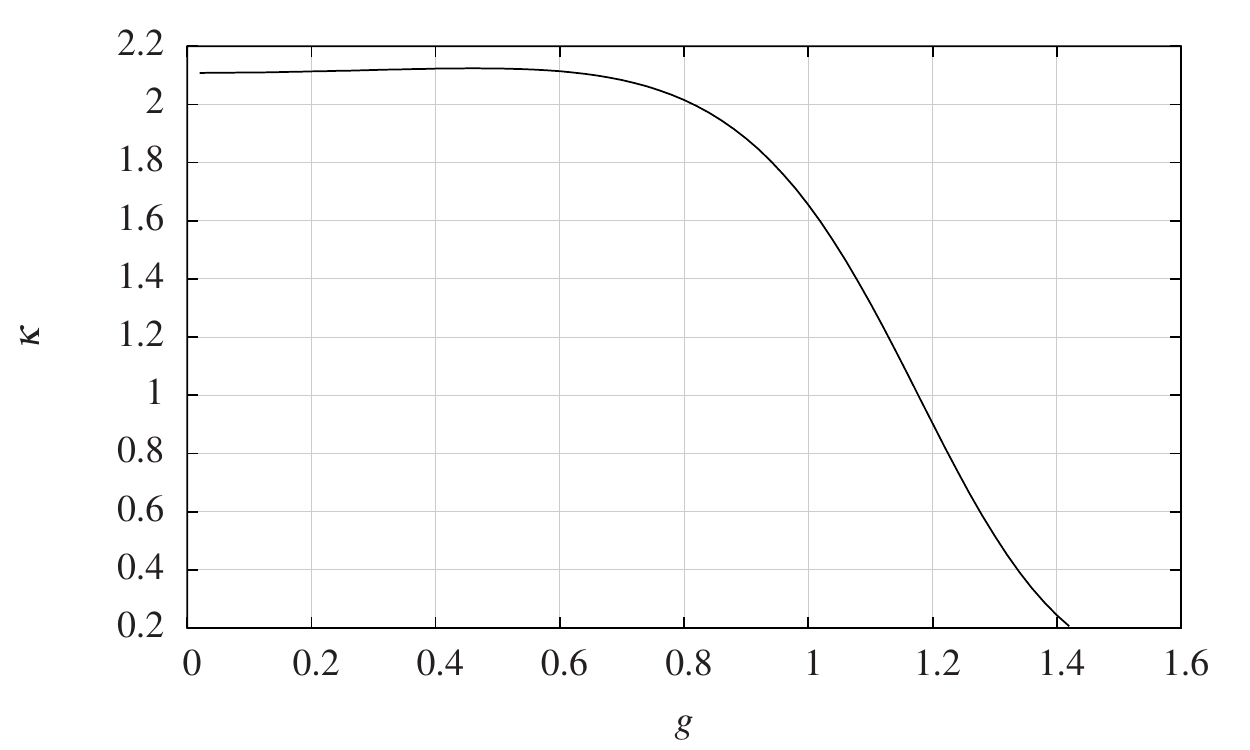}
\includegraphics[height=4.9cm]{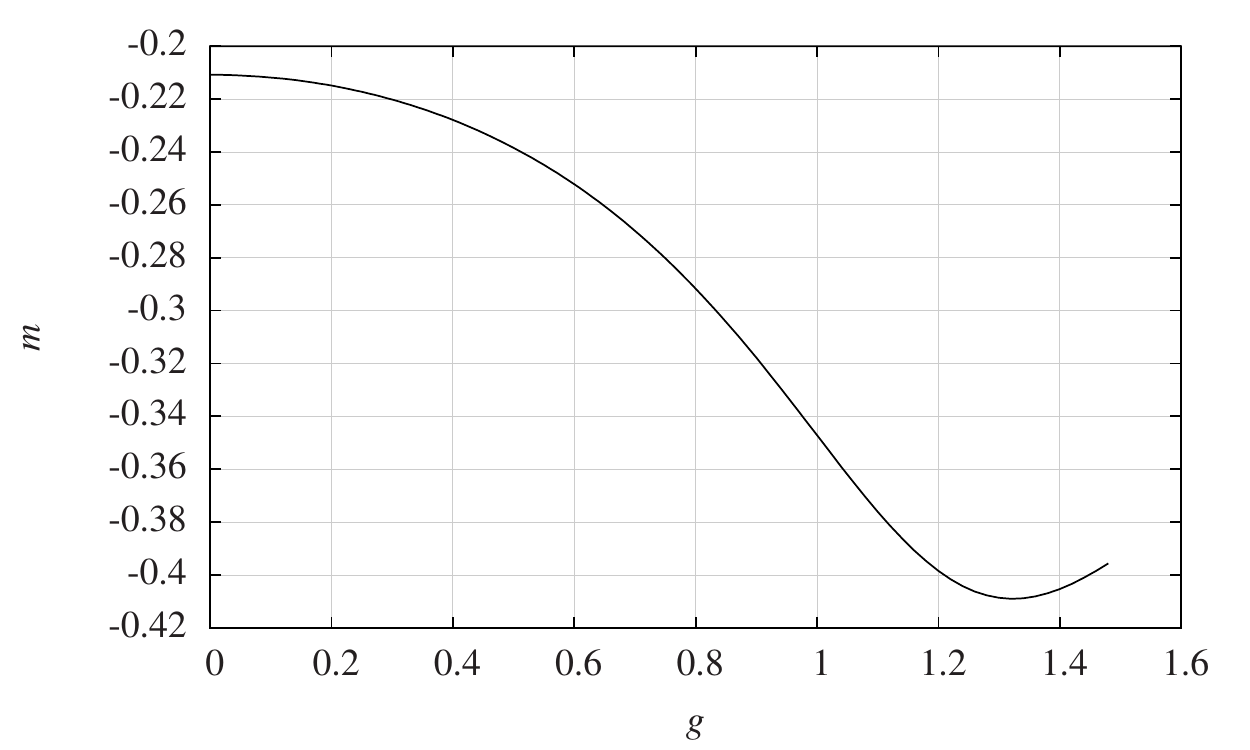}
\caption{The magnetic compression and magnetization density at $H=0$ as a function of $g$.}
\end{figure}
\begin{figure}
\includegraphics[height=6.4cm]{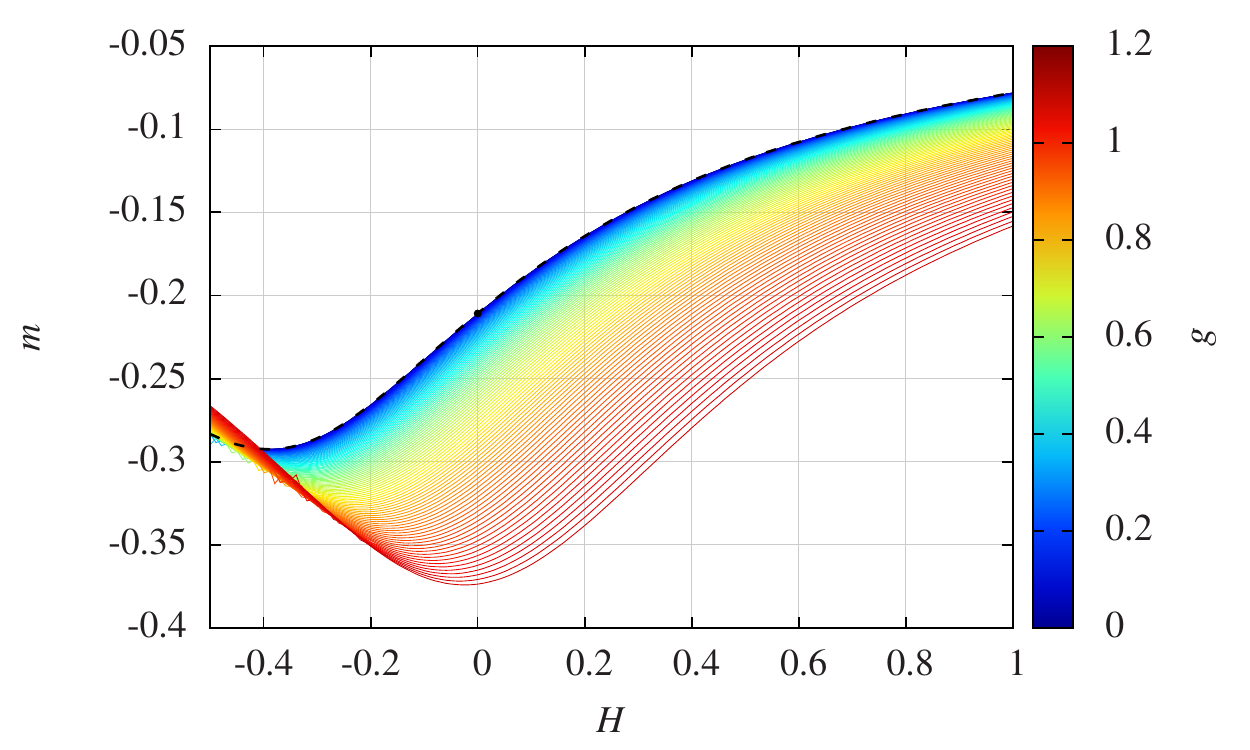}
\caption{The magnetization density as a function of $H$ and $g$.}
\end{figure}
\begin{figure}
\includegraphics[height=6.4cm]{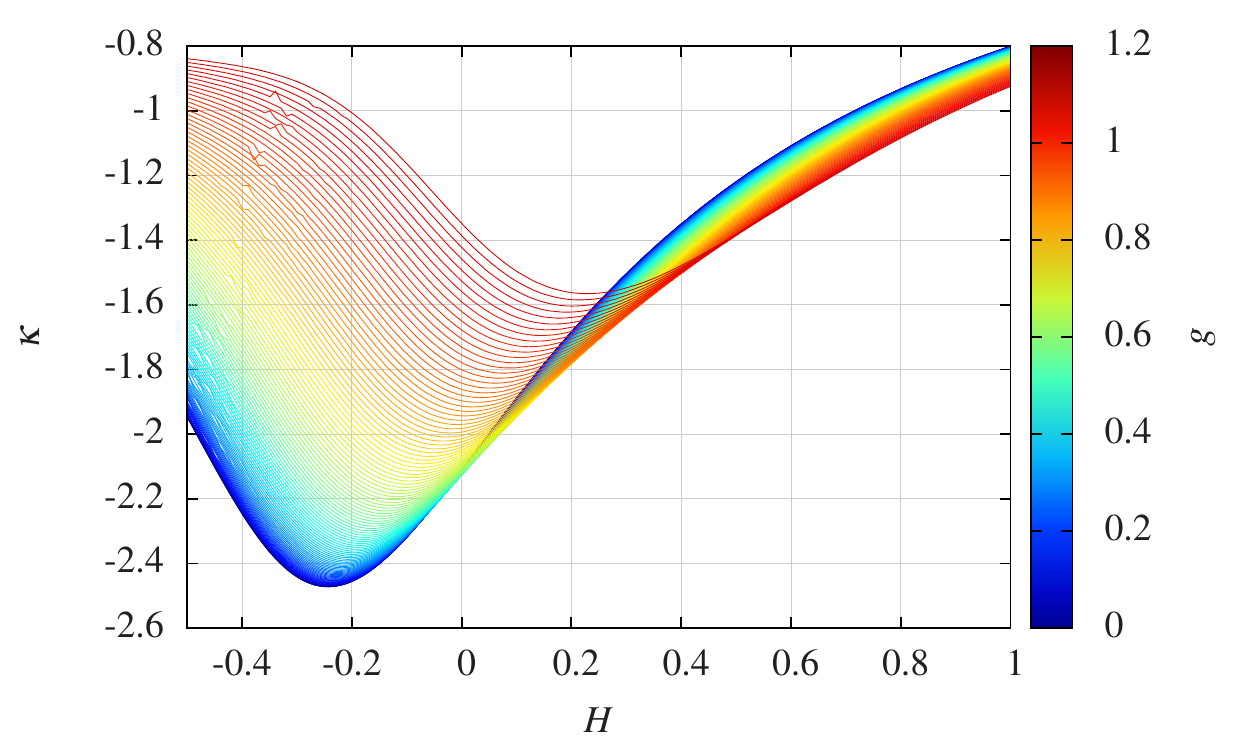}
\caption{The magnetic compressibility as a function of $H$ and $g$.}
\end{figure}
\begin{figure}
\includegraphics[height=6.4cm]{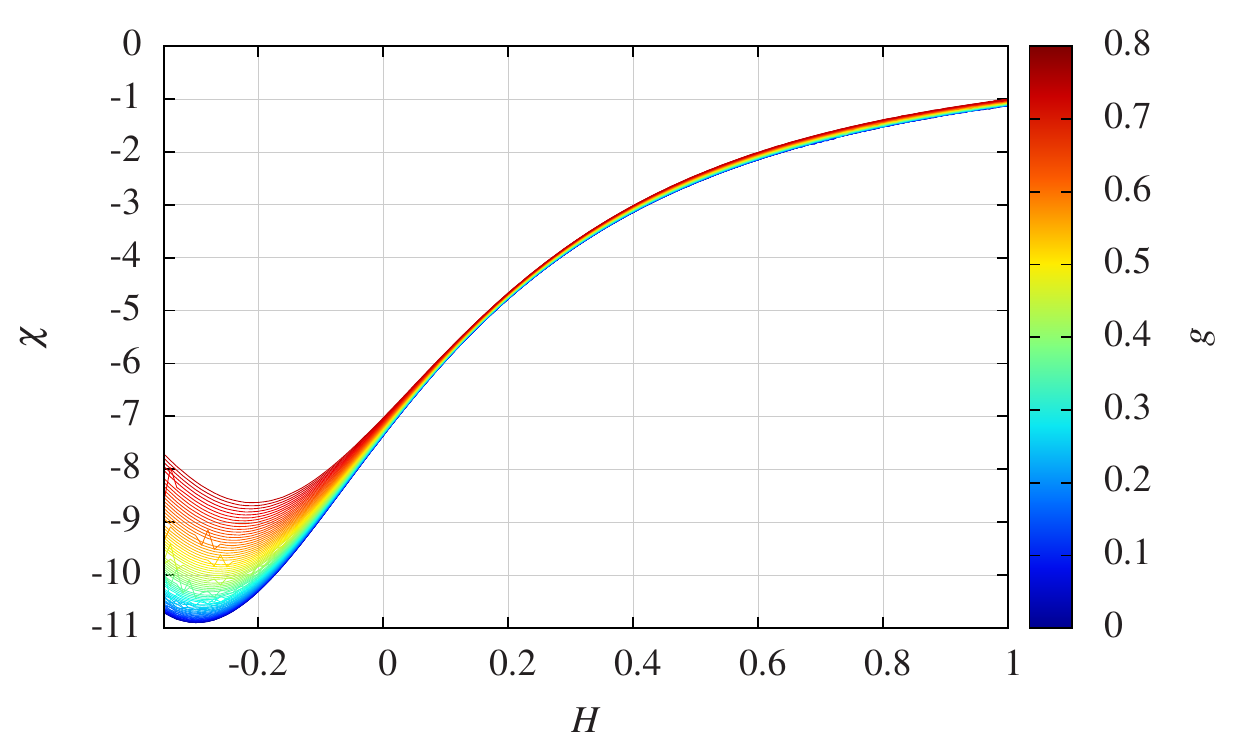}
\caption{The magnetic susceptibility as a function of $H$ and $g$.}
\end{figure}
%%%%%%%%%%%%%%%%%%%%%%%%%%%%%%%%%%%%%%%%%
\subsection{ Non-dynamical constant magnetic field}
%%%%%%%%%%%%%%%%%%%%%%%%%%%%%%%%%%%%%%%%%
Although the system can be reduced to BPS first order equations it is still too complicated to find analytical solutions. However, one may consider a simplified case where the magnetic field is treated as an external field $B=H=\mbox{const}.$. That is to say, we do not consider the back reaction of the system on the magnetic field in the vicinity of the BPS baby skyrmion.  It has been found, after comparison with the numerical results, that this approximation works quite well and provides an exact description in the small electrodynamical coupling constant limit $g \rightarrow 0$. 
%%%%%%%%%%%%%%%%%%%%%%%%%%%%%%%%%%%%%%%%%
\subsubsection{Equation of state $V=V(H)$ and $E=E(H)$ }
%%%%%%%%%%%%%%%%%%%%%%%%%%%%%%%%%%%%%%%%%
As the magnetic field is only a non-dynamical external field, we may reduce the system to one equation where the magnetic field plays the role of a "deformed metric" in which baby skyrmions exist. (In fact, curved metrics may arise in some gravitational context \cite{grav}, which points to another possible application of the BPS skyrmions.) Hence, 
\begin{equation} \label{h-eq-ext}
\sin f \left\{ \partial_y \left[ h_y (1+a)^2 \right] - \frac{\mu^2}{4n^2 \lambda^2} U_h \right\}=0
\end{equation} 
where 
\begin{equation}
B\equiv H = const \;\; \Rightarrow \;\; a= \frac{H r^2}{2n} \;\; \Rightarrow \;\; a = \frac{H}{n} y \equiv \beta y .
\end{equation}
The resulting equation can be analytically solved for the old baby potential
\begin{equation}
U=2h.
\end{equation}
Then,
\begin{equation}
 \partial_y \left[ h_y (1+a)^2 \right] = \frac{\mu^2}{2n^2 \lambda^2} \;\; \Rightarrow \;\;  h_y (1+\beta y)^2 = \frac{\mu^2}{2n^2 \lambda^2} (y-y_0) \label{hy}
\end{equation}
Hence,
\begin{equation}
h(y)=\frac{\mu^2}{2n^2 \lambda^2} \int dy \frac{y-y_0}{(1+\beta y)^2} +{\rm const.}
\end{equation}
with the boundary conditions
\begin{equation}
h(0)=1, \;\; h(y_0)=0, \;\;\; h'(y_0)=0
\end{equation}
where $y_0$ can be finite (compacton) or infinite (usual soliton). However, infinite $y_0$ is excluded by the asymptotic behavior of equation (\ref{hy}). Indeed, for large $y$ we get that $h \sim \ln y$ which contradicts the boundary value for $h$ at infinity. The final solution is
\begin{equation}
h(y) = \left\{
 \begin{array}{cc}
 \frac{\mu^2}{2 n^2\lambda^2 \beta^2} \left[ \frac{1+\beta y_0}{1+\beta y}-\ln \left( \frac{1+\beta y_0}{1+\beta y} \right)  -1 \right] & y \leq y_0 \\
 0 & y\geq y_0
 \end{array} \right.
\end{equation}
where
\begin{equation}
\beta y_0-\ln (1+\beta y_0) = \frac{2n^2\lambda^2 \beta^2}{\mu^2} \label{y0}
\end{equation}
is an equation fixing the size of the compacton. It provides an approximate but exact relation between the two-dimensional "volume" $V=2\pi y_0$ and the external magnetic field 
\begin{equation}
\frac{HV}{2\pi n} - \ln \left( 1+\frac{HV}{2\pi n} \right) = \frac{2\lambda^2 H^2}{\mu^2} .
\label{VHapr}
\end{equation}
The validity of this approximation is restricted by the following condition
\begin{equation}
\frac{g^2\mu^4}{n^4\lambda^2}y_0^2 <<1
\end{equation}
which follows from the equation of motion for the magnetic field when the approximated (non-back reaction) solution is inserted. 
For small magnetic field $\beta y_0 << 1$ we may  use
\begin{equation}
\ln (1+x)= x - \frac{1}{2} x^2 +...
\end{equation}
and then
\begin{equation}
 y_0 = \frac{2n\lambda }{\mu} \;\;\; \Rightarrow \;\;\; V[H=0]=\frac{4\pi \lambda n}{\mu}
\end{equation}
which agrees with the size of the non-gauged case. For large magnetic field we can use $\beta y_0 >> \ln (1+\beta y_0)$. Thus,
\begin{equation}
y_0 = \frac{2n^2\lambda^2 }{\mu^2} \beta \;\;\; \Rightarrow \;\;\;  V =  \frac{4 \pi \lambda^2 n  }{\mu^2} H
\end{equation}
i.e., the size of the solution grows linearly with the magnetic field. 
\\
Next, we consider the energy 
\begin{equation}
E= 2\pi \int_0^{y_0} dy \; 2\lambda^2n^2 (1+a)^2h_y^2 + 2\mu^2 h
\end{equation}
\begin{equation}
= 2\pi \frac{\mu^4}{n^2\lambda^2 }  \int_0^{y_0} dy \; \frac{1}{2} \frac{(y-y_0)^2}{(1+\beta y)^2} + \frac{1}{\beta^2}  \left[ \frac{1+\beta y_0}{1+\beta y}-\ln \left( \frac{1+\beta y_0}{1+\beta y} \right)  -1 \right] 
\end{equation}
\begin{equation}
= 4\pi \frac{\mu^2}{\beta} \left[ \frac{\mu^2}{4\lambda^2 n^2} y_0^2-1 \right] \equiv 4\pi \frac{\mu^2}{\beta} \left[ \frac{y_0^2}{2C} -1 \right] 
\end{equation}
where $C=\frac{2n^2\lambda^2}{\mu^2}$. Hence, we find the relation between the total energy and the external magnetic field, however, in an implicit way
\begin{equation}
E=\frac{4\pi \mu^2 n}{H} \left[ \left( \frac{\mu V}{4\pi  \lambda n} \right)^2 -1\right].
\label{EHapr}
\end{equation}
Equation (\ref{VHapr}) and the last expression are the main results of this section since they provide exact formulas for the $V=V(H)$ and $E=E(H)$ relations in the BPS gauged baby model. 
%%%%%%%%%%%%%%%%%%%%%%%%%%%%%%%%%%%%%%%%%
\subsubsection{Magnetic compressibility}
%%%%%%%%%%%%%%%%%%%%%%%%%%%%%%%%%%%%%%%%%
For small magnetic field $y_0^2 \rightarrow 2C$ and the last expression can be computed using the L'Hospital formula 
\begin{equation}
E[H=0] = \lim_{\beta \rightarrow 0} 4\pi \frac{\mu^2}{\beta} \left[ \frac{y_0^2}{2C} -1 \right] =4 \pi \mu^2 \lim_{\beta \rightarrow 0} \frac{2y_0y'_0}{2C} .
\end{equation}
In order to find $y'_0$ at vanishing $\beta$ we differentiate (\ref{y0})
\begin{equation}
\frac{y_0}{\beta}-\frac{\ln(1+\beta y_0)}{\beta^2} =C .
\end{equation}
Then,
\begin{equation}
y_0^2+\beta y_0y_0'=2C (1+\beta y_0) .
\end{equation}
Now, assuming $y_0=\sqrt{2C} + A \beta$ we find that $A=\frac{2}{3}C$ i.e.,
\begin{equation}
y_0'(\beta=0)=\frac{2}{3} C \label{y0'} .
\end{equation}
We plot the numerical results for $y_0 (H=0)$ and $y'_0 (H=0)$ for general coupling $g$ (i.e., with the backreaction taken into account) in Fig. 5.

Then the energy is
\begin{equation}
E[H=0] =  \frac{16\pi}{3} \mu \lambda n
\end{equation}
which agrees with the non-gauged case. On the other hand, for large value of the magnetic field we find that
\be
E=4\pi \lambda^2 n H .
\ee
Another consequence of (\ref{y0'}) is that the magnetic compressibility is finite
\begin{equation}
\left. \kappa_{mag}^0  \equiv \frac{1}{V} \frac{\partial V}{\partial H} \right|_{H=0}= \frac{2 \lambda}{3\mu} 
\end{equation}
It is quite interesting that the magnetic compressibility very weakly depends on the electromagnetic coupling constant for a wide range of $g$. In fact, $\kappa_{mag} \approx \kappa_{mag} (g=0) = 2.1082$ for $g \in [0,0.7]$, see Fig. 6. Hence, the non-backreaction approximation works especially well for the magnetic compressibility.  
\\
Moreover, we can also obtain the magnetic compressibility for large magnetic field. Now,
\be
\kappa_{mag} (H) \sim \frac{1}{H} .
\ee
Hence, asymptotically the magnetic compressibility tends to zero.  
%%%%%%%%%%%%%%%%%%%%%%%%%%%%%%%%%%%%%%%%%
\subsubsection{Magnetization and ferromagnetic medium}
%%%%%%%%%%%%%%%%%%%%%%%%%%%%%%%%%%%%%%%%%
Another interesting quantity is the magnetization at vanishing external field, $M^0=-
\left. \frac{\partial E}{\partial H} \right|_{H=0}.
$
Then,
\begin{equation}
\left. \frac{\partial E}{\partial H} \right|_{H=0} =\frac{1}{n} \left. \frac{\partial E}{\partial \beta} \right|_{\beta=0} = \frac{4\pi \mu^2}{n\beta^2}\left.  \left( 1- \frac{y_0^2}{2C} +\frac{2y_0y_0' \beta}{2C} \right)  \right|_{\beta=0} .
\end{equation}
Hence,
\begin{equation}
\left. \frac{\partial E}{\partial H} \right|_{H=0} =\left.  \frac{4\pi \mu^2}{n2\beta} \left(-\frac{2y_0y_0'}{2C}+\frac{2y_0y_0'}{2C}+\frac{2y_0'^2 \beta}{2C}+\frac{2y_0y_0''\beta}{2C} \right) \right|_{\beta=0} =\left.  \frac{4\pi \mu^2}{2C} (y_0'^2+y_0y_0'') \right|_{\beta=0} 
\end{equation}
Again, from (\ref{y0}) we find that 
\begin{equation}
y''_0(\beta =0) = (2C)^{3/2} \frac{1}{18} 
\end{equation}
and
\begin{equation}
\left. \frac{\partial E}{\partial H} \right|_{H=0} = 4\pi \frac{2}{3} \lambda^2 n .
\end{equation} 
Then, we can find the magnetization in the vicinity of the vanishing magnetic field
\be
M^0=-\left. \frac{\partial E}{\partial H} \right|_{H=0} = -  4\pi \frac{2}{3} \lambda^2 n
\ee
and the magnetization density
\be
m^0=- \frac{1}{V} \left. \frac{\partial E}{\partial H} \right|_{H=0} = - \frac{2}{3} \lambda \mu
\ee
which is negative for the baby skyrmions (remember $n>0$).
For general coupling $g$ (with the back reaction included) we plot the magnetization density in Fig. 6.

Due to the nonlinearity of the model, the magnetization is not $H$-independent. In fact, for a big enough value of the magnetic field we get 
\be
M (H)  = -4\pi \lambda^2 n
\ee 
and therefore the magnetization density goes to 0 as $1/H$. These exact results find a perfect agreement with the numerical computation. 
\begin{table}
\begin{center}
\begin{tabular}{|c|c|c|c|c|}
\hline
$E^0$ & $V^0$ & $\kappa_{mag}^0$ &  $m^0$&  $\chi^0_d$ \\
\hline 
\hline
& & & & \\
$\; \frac{16\pi}{3} \mu \lambda n \; $ & $\; 4\pi \frac{\lambda}{\mu}n \;$  & $\;\frac{2}{3} \frac{\lambda}{ \mu}\;$ &$\;-\frac{2}{3}\lambda \mu \;$& $-\; \frac{8}{45}\lambda^2\; $ \\
& & & & \\
%\hspace*{0.1cm}
\hline
\end{tabular}  
\caption{Energy, volume, magnetic compressibility, magnetization density and density of the magnetic susceptibility for the non-back reaction approximation at $H=0$. } \label{table}
\end{center}
\end{table}
%\\
%Indeed, the external magnetic field is always repealed from the BPS gauged baby skyrmions. 
\\
Another quantity relevant for the study of magnetic properties of a medium is the magnetic susceptibility defined as
\be
\chi =  \frac{\partial M }{\partial H}= - \frac{\partial^2 E}{\partial H^2}
\ee
Then using the equation of state for the energy we find that at $H=0$
\be
 \left. \chi^0 = -\frac{1}{n^2} \frac{\partial^2 E}{\partial \beta^2} \right|_{\beta=0} = -\frac{4\pi \mu^2}{n^2} \left[ \frac{2}{\beta^3} \left( \frac{y_0^2}{2C} -1\right) -\frac{2y_0y'_0}{C\beta^2} + \frac{y_0'^2}{C\beta} +\frac{y_0y_0''}{C\beta} \right]_{\beta=0} 
 \ee
\be
= -\frac{4\pi \mu^2}{3C} [3y_0'y_0''+y_0y_0''']_{\beta=0}
\ee
Now, from the volume-magnetic field equation of state we get that
\be
y_0'''(\beta=0)= - \frac{(2C)^2}{45}
\ee
Then the final result for the magnetic susceptibility at $H=0$ is
\be
\chi^0 = - \frac{32\pi}{45} \frac{\lambda^3}{\mu} n
\ee
and its density
\be
\left. \chi^0_d= -\frac{1}{V} \frac{\partial^2 E}{\partial H^2} \right|_{\beta=0} = -  \frac{8}{45} \lambda^2
\ee
which are negative for any values of the parameters of the model. The exact analytical result is confirmed by numerical computations. For higher values of the magnetic field the susceptibility tends to zero. 

\vspace*{0.2cm}

Let us now interpret the results obtained above. First of all, as we know from \cite{BPS-g}, the gauged BPS baby skyrmions always possess a non-zero flux of the magnetic field - even without external magnetic field, i.e., for the boundary condition $H=0$. In other words, after gauging the BPS Skyrme model there are no topological solitons without magnetic field. Hence, the BPS skyrmions are like two dimensional magnets with a permanent magnetization. Such magnets behave as ferromagnets since they add positively, i.e., the total magnetic flux of a baryon number $n$ baby skyrmion is $n$ times the flux of a $n=1$ soliton. 
\\
Interestingly enough, one can make the magnetic susceptibility arbitrarily small. 
\\
Observe that the response to the external magnetic field is the standard one, in the sense that the size of the compacton as well as the energy have a finite first (and higher) derivative. Finally, we plot the numerical results for the magnetization density, the magnetic compressibility and the magnetic susceptibility in Figs. 7-9.
%%%%%%%%%%%%%%%%%%%%%%%%%%%%%%%%%%%%%%%%%
\section{Pressure}
%%%%%%%%%%%%%%%%%%%%%%%%%%%%%%%%%%%%%%%%%
%%%%%%%%%%%%%%%%%%%%%%%%%%%%%%%%%%%%%%%%%
\subsection{Pressure in the ungauged BPS baby Skyrme model}
%%%%%%%%%%%%%%%%%%%%%%%%%%%%%%%%%%%%%%%%%
There is a natural way to introduce pressure in the BPS (baby) Skyrme model, for details we refer to \cite{pres}.  Let us first rewrite the BPS baby Skyrme model as
\be
\mathcal{L}=-\frac{\lambda^2}{8} j_\mu^2 - \mu^2 U
\ee
where
\be
j_\mu = \epsilon_{\mu \nu \rho} \vec{\phi} \cdot (\partial^\nu \vec{\phi} \times \partial^\rho \vec{\phi})
\ee
is the topological current and 
\be
j_0=2 \;q, \;\;\;\; q\equiv \phi \cdot (\partial_1\vec{\phi} \times \partial_2\vec{\phi} ) .
\ee
Then, for static configurations, the components of the energy-momentum tensor are
\be
T^{00}=\frac{\lambda^2}{8} j_0^2 + \mu^2 U=\mathcal{E}, \;\;\;\;\;\;\;\; T^{ij} = \delta^{ij} \left(\frac{\lambda^2}{8} j_0^2 - \mu^2 U \right) \equiv \delta^{ij} \mathcal{P}
\ee
where $\mathcal{E}, \mathcal{P}$ are the energy density and the pressure. Obviously, for zero pressure we obtain the BPS equation for the (ungauged) BPS baby Skyrme model. In fact, BPS equations are often referred to as zero pressure conditions \cite{bazeia}. However, it is a matter of fact that equation
\be
\frac{\lambda^2}{8} j_0^2 - \mu^2 U =P
\ee
with a constant value of the pressure $\mathcal{P}=P$ is a first integral of the full static equations of motion \cite{pres}, where the pressure is now an integration constant. Hence, we find a one-parameter set of first order equations which correspond to different fixed values of the pressure. 
\\
In the case of the old baby potential $(U=2h)$ we get
\begin{equation}
\lambda^2 n^2 h_y^2 - \mu^2 h=0
\end{equation}
which can be easily generalized to non-zero pressure
\begin{equation}
2\lambda^2 n^2 h_y^2 -2 \mu^2 h=P.
\end{equation}
Hence,
\begin{equation}
\lambda n h_y = - \mu \sqrt{h + \tilde P}
\end{equation}
where $\tilde P = P/2\mu^2$.
It is convenient to introduce $z\equiv \frac{\mu}{n\lambda} y$ and then
\begin{equation}
h_z = - \sqrt{h + \tilde P}
\end{equation}
with the conditions $h(0)=1$ and $h(Z)=0$, where $Z$ is the size of the compacton in the presence of the external pressure. We find
\begin{equation}
h=\frac{1}{4} (z-z_0)^2-\tilde P, \;\;\; z \leq Z
\end{equation}
where 
\begin{equation}
z_0=2\sqrt{1+\tilde P}, \;\;\;\; Z=z_0-2\sqrt{\tilde P} = 2 ( \sqrt{1+\tilde P} - \sqrt{\tilde P} )
\end{equation}
Hence, the volume-pressure equation of state is
\begin{equation}
V=\pi R^2 = \pi \frac{2\lambda n }{\mu} Z = \frac{4\pi \lambda n}{\mu} (\sqrt{1+\tilde P} - \sqrt{\tilde P}) .
\end{equation}
Similarly, one can compute the energy 
\be
E=4\pi \mu \lambda |n| \left[ \frac{4}{3}(1+\tilde{P})^{3/2}+\frac{2}{3}\tilde{P}^{3/2} -2\tilde{P} \sqrt{1+\tilde{P}} \right] .
\ee
Observe, that the energy has a smooth first derivative w.r.t. to the pressure, while the corresponding derivative of the volume diverges, corresponding to an infinite (isothermal) compressibility, \cite{pres}. 
\\
Another example is the new baby potential $V=2h(1-h)$. Then, the non-zero pressure equation
\be
h_z=-\sqrt{h(1-h)+\tilde P}
\ee
gives
\be
\frac{1-2h}{2\sqrt{\tilde P +h(1-h)}}= \tan (z-z_0), \;\;\;\;\; z\leq Z
\ee
where 
\be
\tan z_0= \frac{1}{2\sqrt{\tilde P}}, \;\;\; Z=2z_0
\ee
Hence,
\be
V=\pi \frac{2\lambda n}{\mu} Z = \frac{4\pi \lambda n}{\mu} \arctan \frac{1}{2\sqrt{\tilde P}}
\ee
%%%%%%%%%%%%%%%%%%%%%%%%%%%%%%%%%%%%%%%%%
\subsection{Pressure in the gauged BPS baby Skyrme model}
%%%%%%%%%%%%%%%%%%%%%%%%%%%%%%%%%%%%%%%%%
The pressure may be introduced in the same manner in the gauged model. The corresponding energy-momentum tensor for static configurations reads
\be
T^{ij}=\frac{1}{2} \left( \lambda^2 Q^2 -2\mu^2U+\frac{1}{g^2} B^2\right) \delta^{ij}
\ee
which still is the energy-momentum tensor of a perfect fluid. Again, the pressure is defined as
\be
T^{ij}=\delta^{ij} \mathcal{P} 
\ee 
and is zero for the BPS solutions.
Quite interestingly there is a generalization of the BPS equations which leads to a non-zero but constant value of the pressure $\mathcal{P}=P$. Namely consider the usual BPS equations
 \begin{equation}
Q= W' \label{bps1-p}
\end{equation}
\begin{equation}
B = - g^2\lambda^2W \label{bps2-p}
\end{equation}
where the superpotential is defined by
\be
\lambda^2W'^2+g^2\lambda^4W^2=2\mu^2U +2P.
\ee
Then, this set of equations again leads to the full e.o.m.  and gives $T^{ij}=P\delta^{ij}$.
\\
It is rather surprising that the pressure may be introduced simply by a small change in the definition of the superpotential $W$. There is also an intriguing similarity between the non-zero pressure configurations and non-extremal solitons in the fake supersymmetric theories. Indeed, the pressure seems to play exactly the same role as the non-extremality parameter \cite{susy}. For example, it modifies the superpotential equation in a very similar manner.  
%%%%%%%%%%%%%%%%%%%%%%%%%%%%%%%%%%%%%%%%%
\subsection{Pressure in the gauged BPS baby Skyrme model with asymptotically constant magnetic field}
%%%%%%%%%%%%%%%%%%%%%%%%%%%%%%%%%%%%%%%%%
In this case we get the BPS equation for the asymptotically constant magnetic field with the superpotential defined as in the upper analyzed non-zero pressure case
\begin{equation}
Q= W \label{bps1-pt}
\end{equation}
\begin{equation}
B = - g^2\lambda^2 W +H \label{bps2-pt}
\end{equation}
and
\be
\lambda^2W'^2+g^2\lambda^4W^2 - 2\lambda^2 WH=2\mu^2U +2P.
\ee
It is a nice feature of the gauged BPS baby Skyrme model that both pressure and asymptotically constant magnetic field may be introduced by modifications of the equation defining the superpotential while the BPS equations remain unchanged. 

%%%%%%%%%%%%%%%%%%%%%%%%%%%%%%%%%%%%%%%%%
\section{Pressure and the old baby potential}
%%%%%%%%%%%%%%%%%%%%%%%%%%%%%%%%%%%%%%%%%
%%%%%%%%%%%%%%%%%%%%%%%%%%%%%%%%%%%%%%%%%
\subsection{Numerical computations}
%%%%%%%%%%%%%%%%%%%%%%%%%%%%%%%%%%%%%%%%%
We solve the non-zero pressure generalized BPS equation for the old baby potential (\ref{old-pot}), with the axially symmetric ansatz
\begin{equation}
2nh_y(1+a)=-\frac{1}{2}W_h
\end{equation}
\begin{equation}
na_y=-g^2\lambda^2 W +H
\end{equation}
and
\be
\frac{\lambda^2}{4}W_h^2+g^2\lambda^4W^2 - 2\lambda^2 WH=4\mu^2h +2P
\ee
with the following boundary condition 
\be
h(0)=1, \;\;\; h(y_P)=0,
\ee
\be
a(0)=0, \;\;\; a_y(y_P)=\frac{H}{n}
\ee
Here, $y_P$ is the size of the compacton for a non-zero value of the pressure $P$.
Again, for numerics we assume $\lambda=1, n=1$ and take $\mu^2=0.1$ and $g=0.1$. The superpotential obeys the boundary conditions
\be
W(h=0)=0, \;\;\; W_h(0)=\frac{2\sqrt{2P}}{\lambda} .
\ee
Now, we find solutions for a few fixed $H$ with different values of the pressure $P$, see Figs.  10, 11.
\begin{figure}
\includegraphics[height=4.9cm]{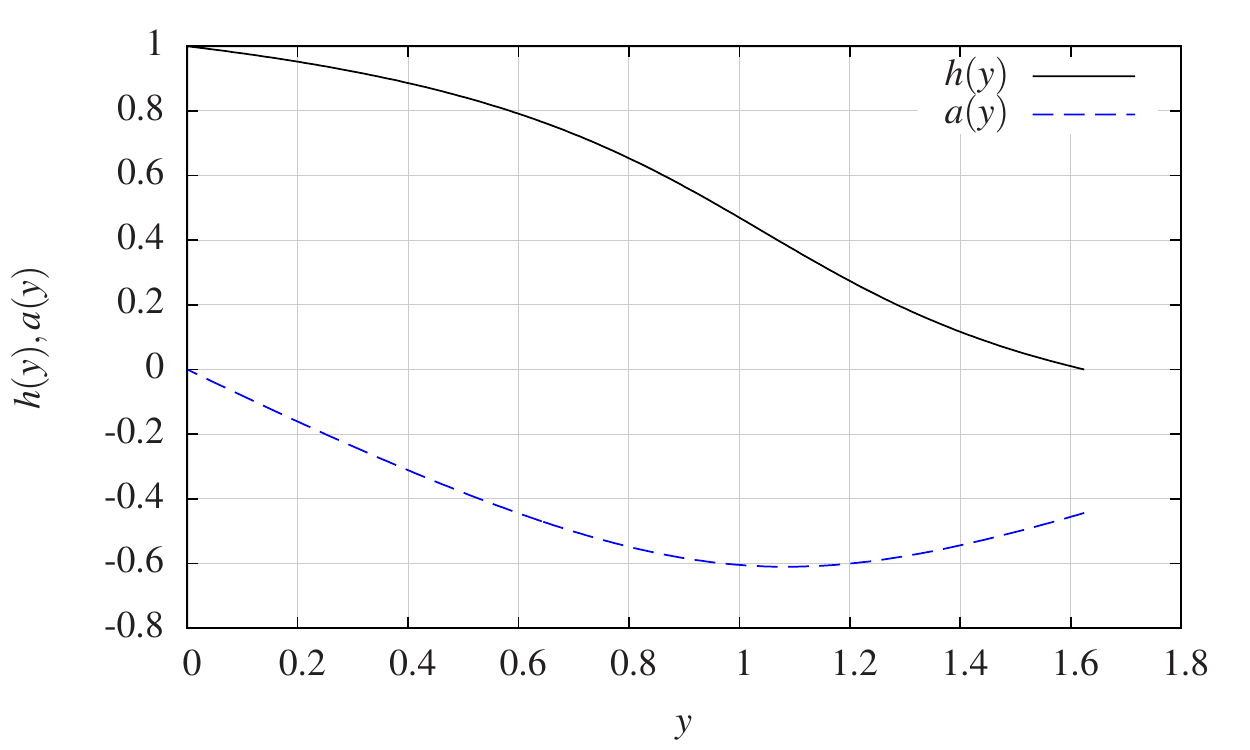}
\includegraphics[height=4.9cm]{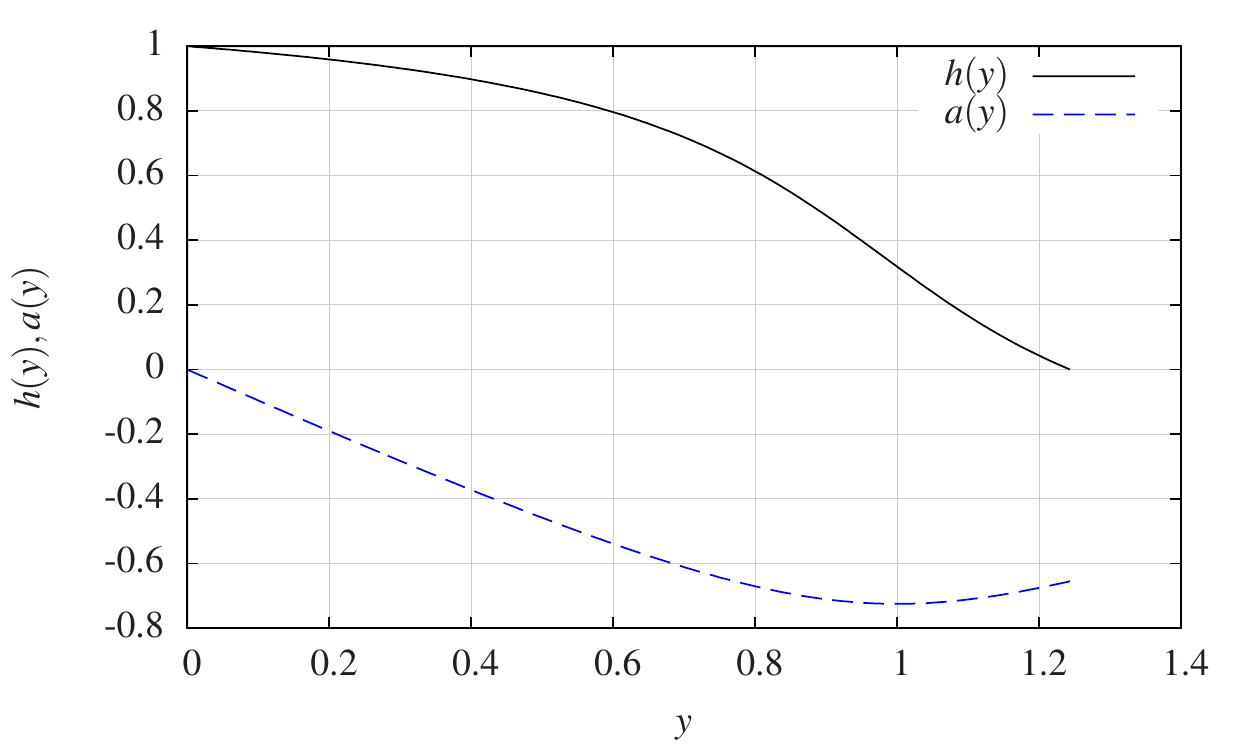}
\includegraphics[height=4.9cm]{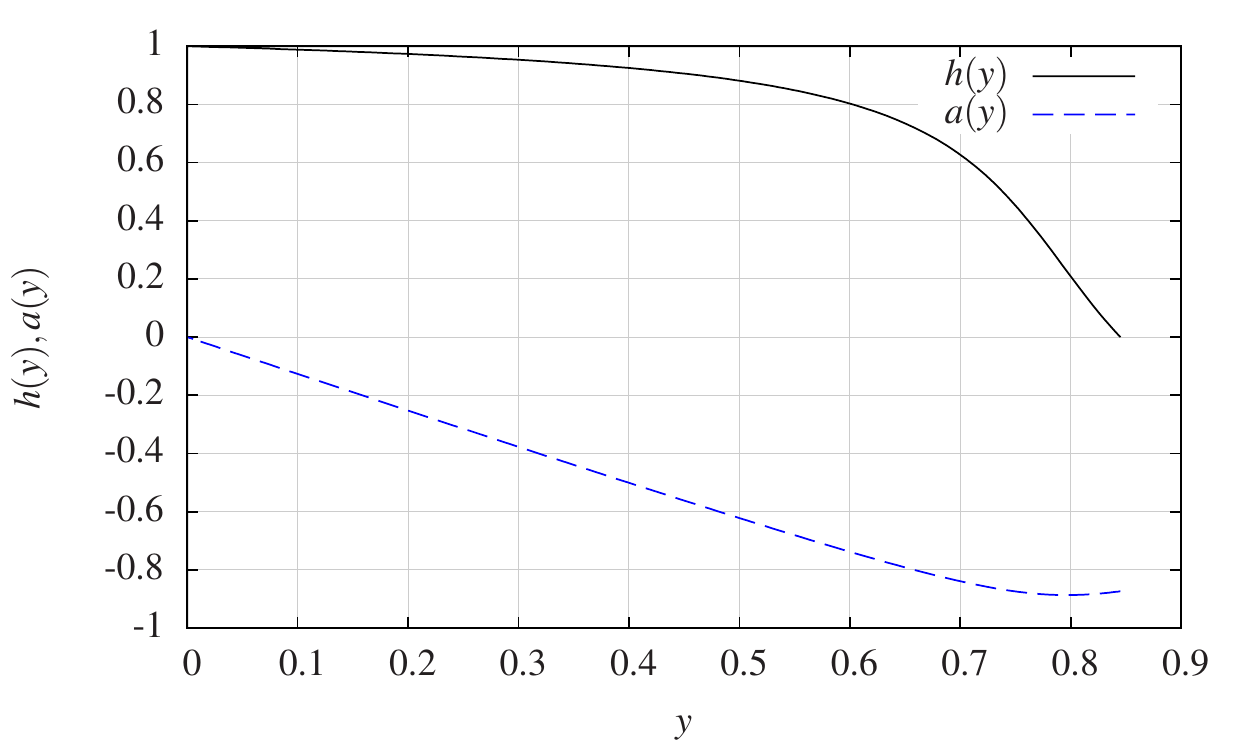}
\includegraphics[height=4.9cm]{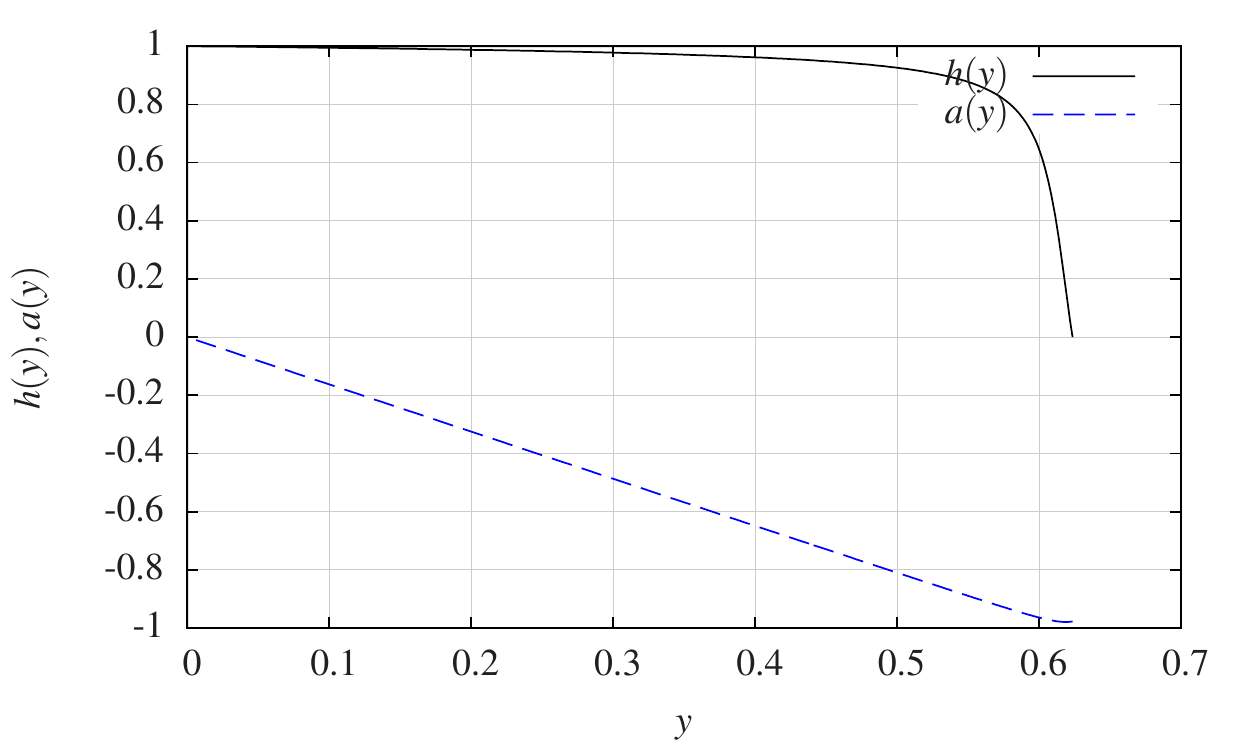}
\caption{The baby skyrmion profile $h$ and magnetic potential $a$ for $H=0.5$ and $P=0.1, 0.2, 0.5, 1$. Here $g=1$.}
\end{figure}
\begin{figure}
\includegraphics[height=4.9cm]{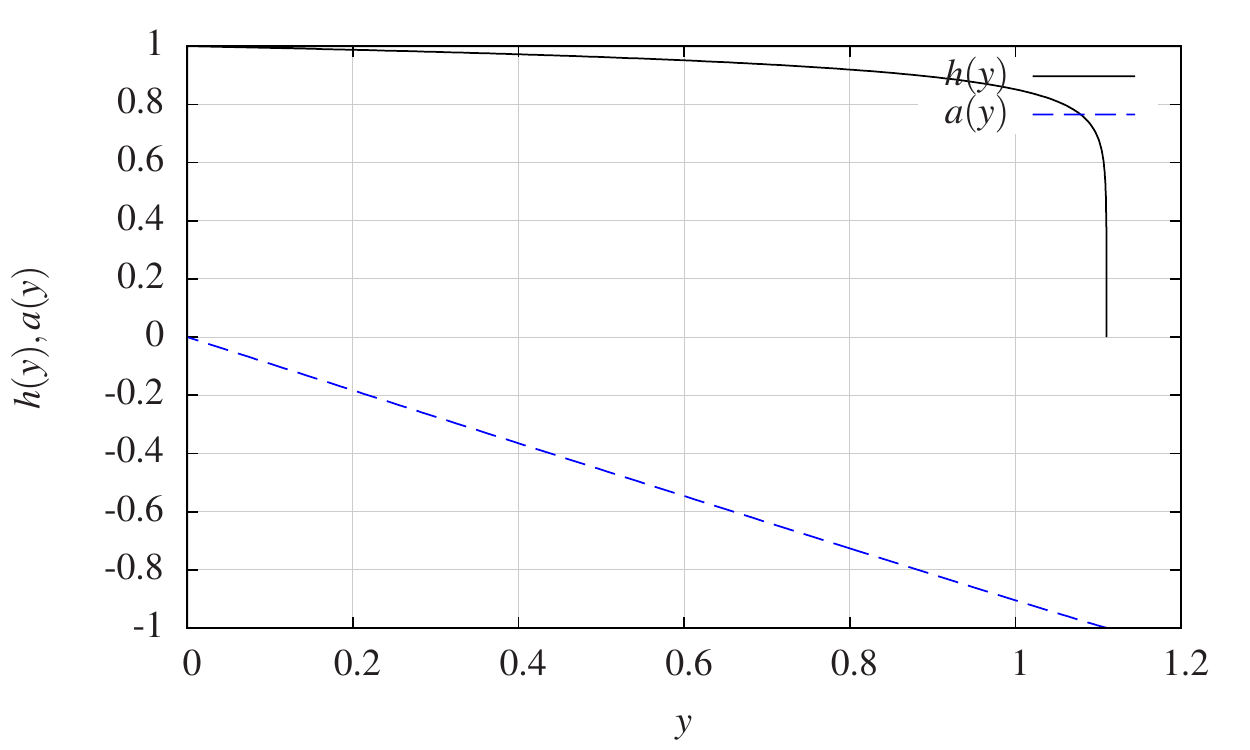}
\includegraphics[height=4.9cm]{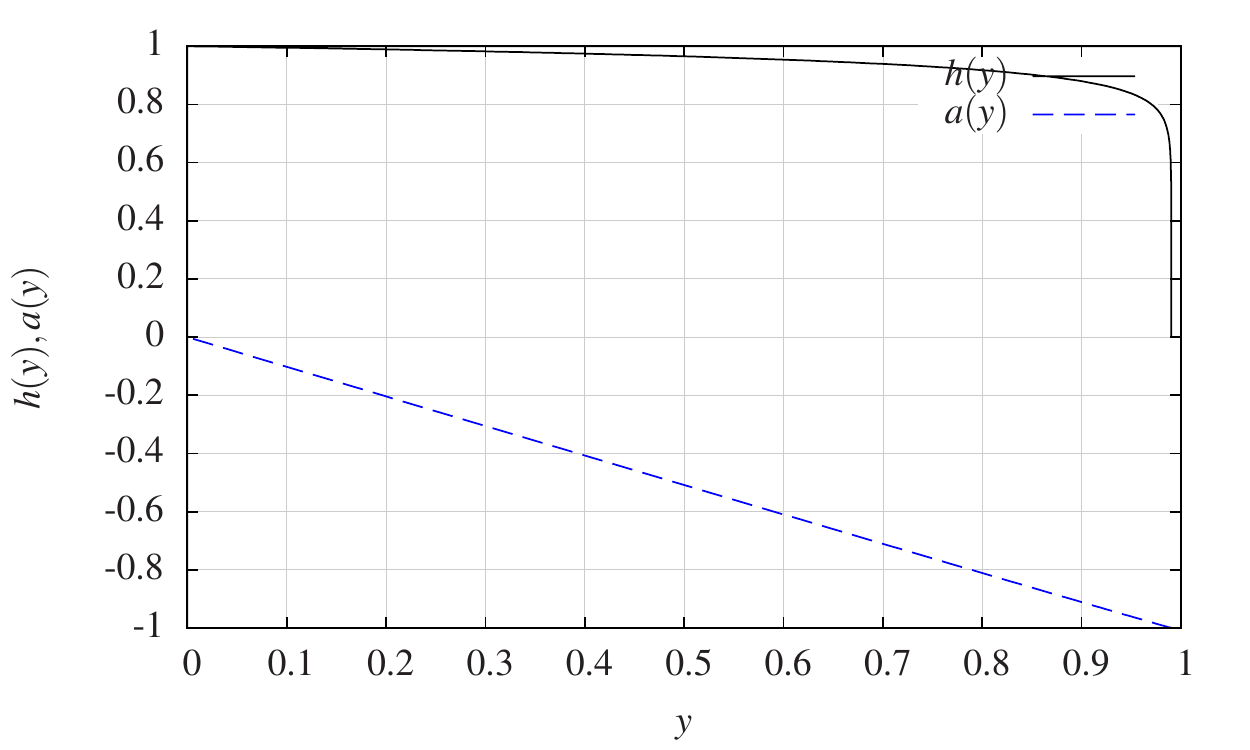}
\includegraphics[height=4.9cm]{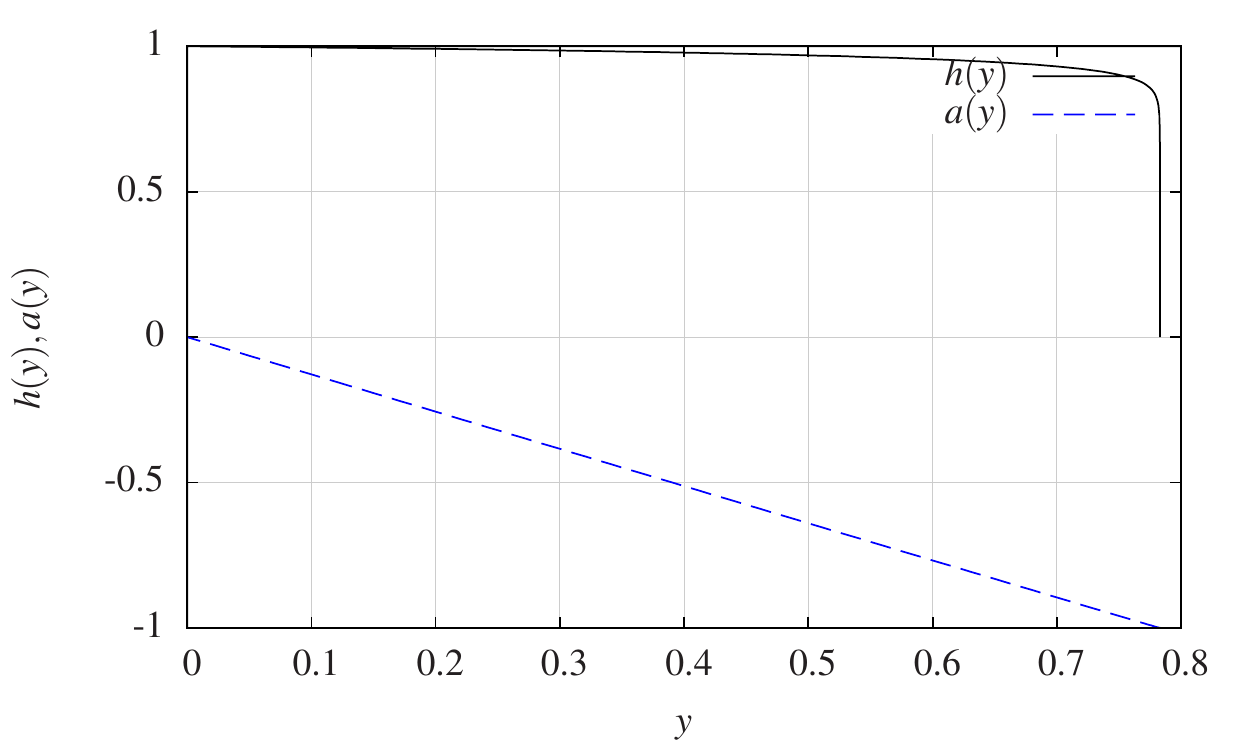}
\includegraphics[height=4.9cm]{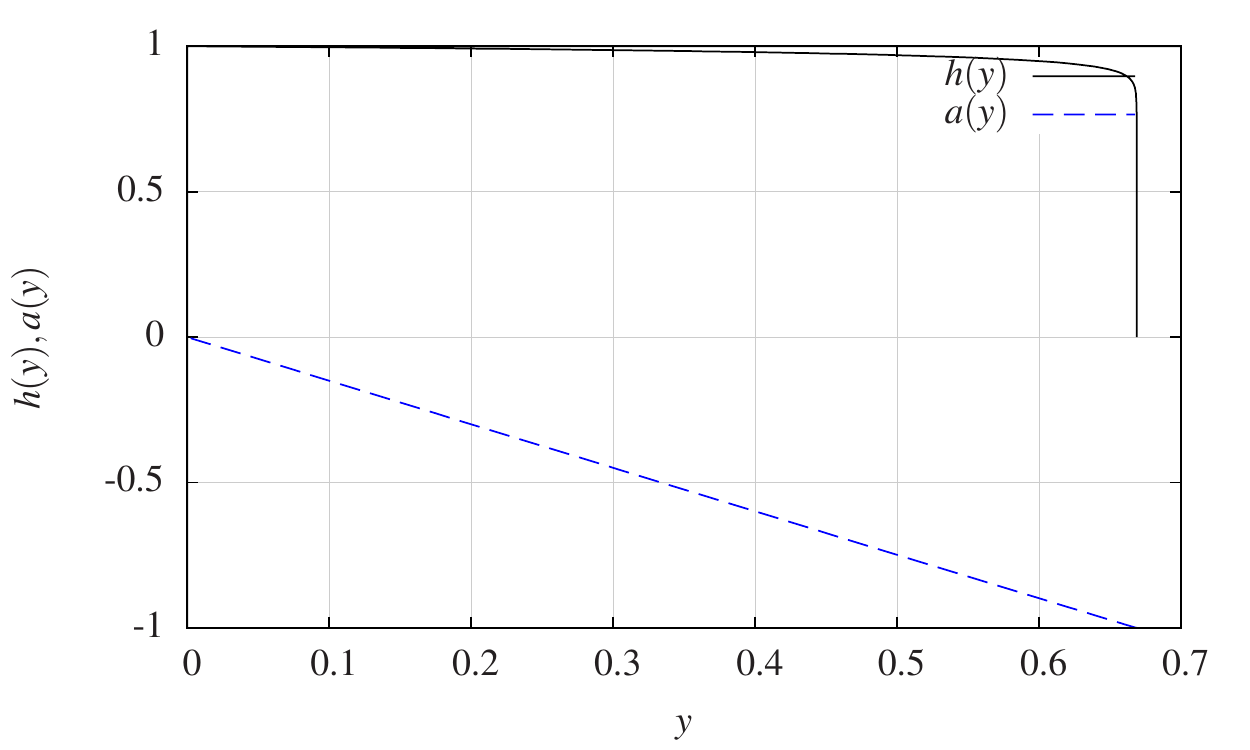}
\caption{The baby skyrmion profile $h$ and magnetic potential $a$ for $H=-0.5$ and $P= 0.1, 0.2, 0.5, 1$. Here $g=1$.}
\end{figure}
\begin{figure}
%\hfill
\includegraphics[height=4.9cm]{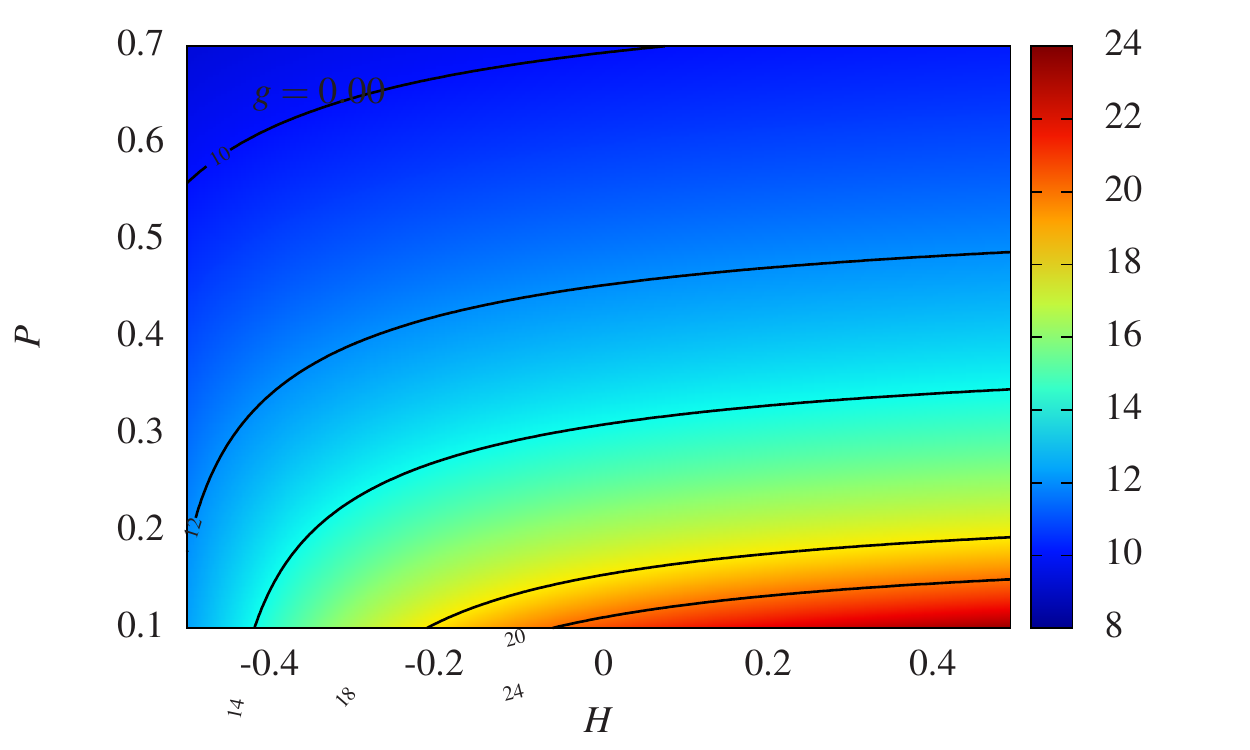}
\includegraphics[height=4.9cm]{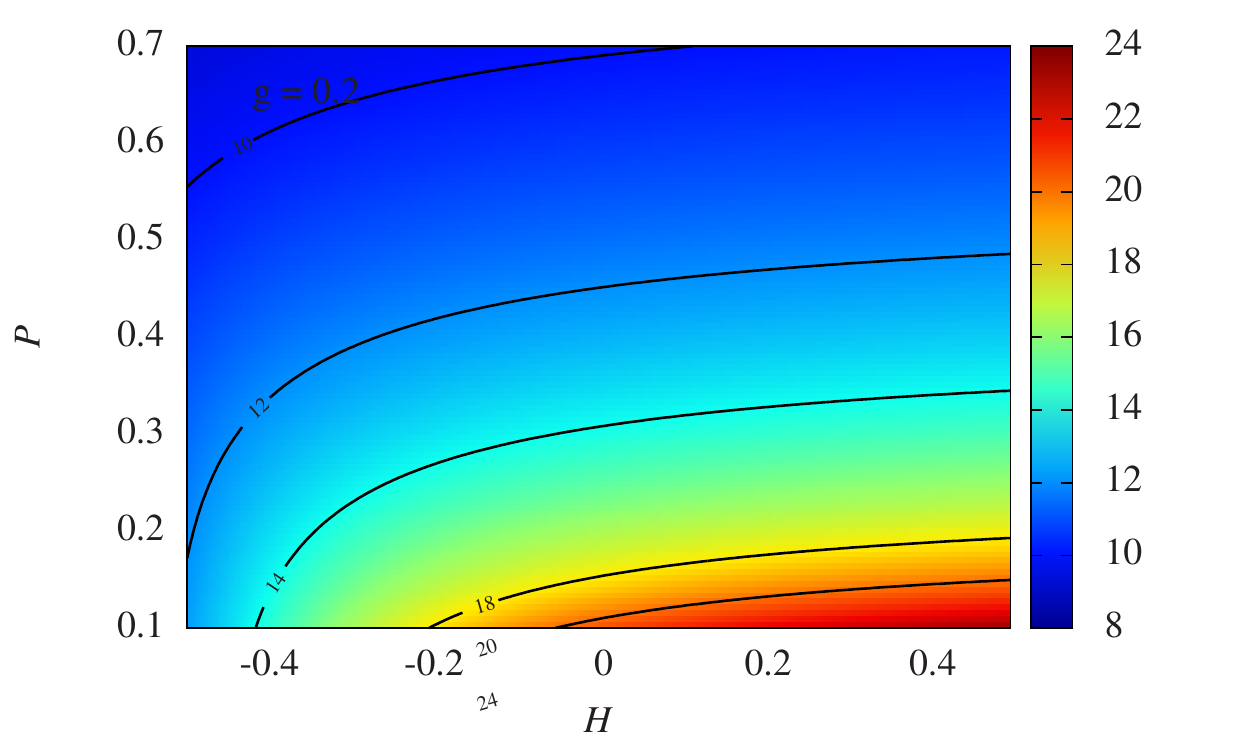}
% \hfill
\caption{The equation of state $V=V(H,P)$ for the non-back reaction approximation (left figure) and for $g=0.2$ (right figure).}
\end{figure}
\begin{figure}
%\hfill
\includegraphics[height=8.6cm]{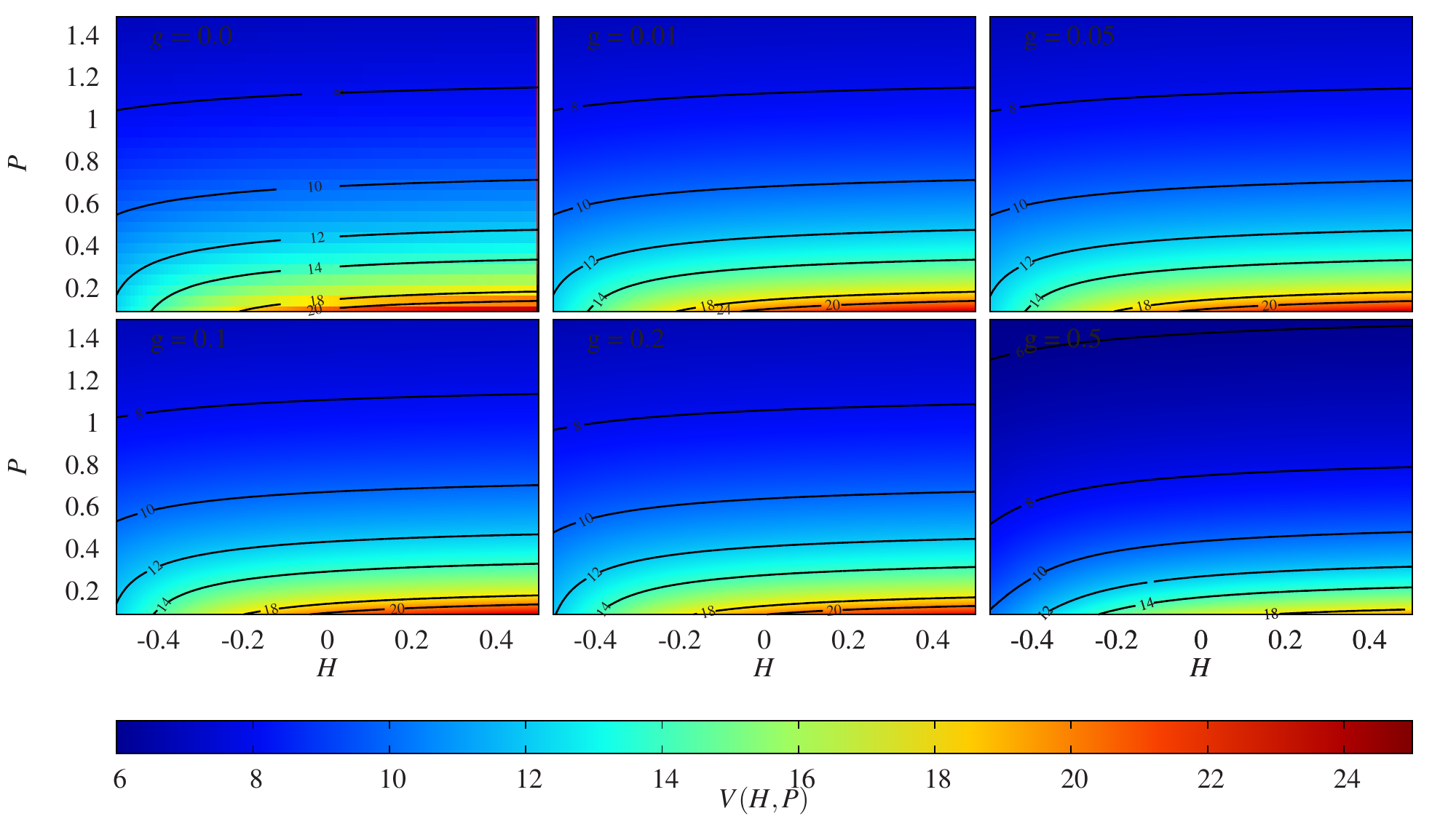}
% \hfill
\caption{The equation of state $V=V(H,P)$ for different values of $g$.}
\end{figure}
%%%%%%%%%%%%%%%%%%%%%%%%%%%%%%%%%%%%%%%%%
\subsection{The boundary pressure approach}
%%%%%%%%%%%%%%%%%%%%%%%%%%%%%%%%%%%%%%%%%
Non-zero pressure requires a solution of the BPS equations with the properly modified superpotential equation which, in general, is a complicated computational problem. However, as we are dealing with BPS models the pressure is constant inside the soliton and can, therefore, also be introduced as a non-zero derivative boundary condition for the matter field. Due to that, we can avoid to solve the superpotential equation. This, together with the non-dynamical magnetic field approximation, (which appeared to be a quite good approximation in the zero-pressure case) will lead us to an approximate but exact expression for the equation of state with non-zero values of $H$ and $P$. 
%%%%%%%%%%%%%%%%%%%%%%%%%%%%%%%%%%%%%%%%%
\subsubsection{The non-gauged BPS baby Skyrme model}
%%%%%%%%%%%%%%%%%%%%%%%%%%%%%%%%%%%%%%%%%
Here we show that the pressure can be introduced by a non-zero value for the derivative of the baby skyrme field at the compacton boundary. In fact, this approach will give {\it exactly} the same equation of state as before. 
\\
Let us again consider the equation of motion of the BPS baby Skyrme model with the old baby potential
\be
4\lambda^2n^2h_{yy}-2\mu^2=0 \;\;\; \Rightarrow \;\;\; h(y)=\frac{\mu^2}{4n^2\lambda^2} (y-y_0)^2+C
\ee
where $y_0, C$ are integration constants. Now, we modify the boundary condition 
\be
h(0)=1, \;\;\; h(y_p)=0, \;\;\; h_y(y_p)=p
\ee
where $y_p$ is the size of the compacton for non-zero value of the derivative $h_y$ at the boundary $y_p$.  Hence, we get a one $p$- parameter family of solutions ($p$ is negative)
\be
h(y)=1+\frac{\mu^2}{4n^2\lambda^2} \left[ y^2 - 2y \left( y_p -\frac{2n^2\lambda^2}{\mu^2} p \right) \right]
\ee
where the compacton radius satisfies
\be
y_p^2-\frac{4n^2\lambda^2}{\mu^2} y_p p - \frac{4n^2\lambda^2}{\mu^2}=0. \label{size-pres}
\ee
It remains to connect the parameter $p$ with the pressure $P$, which is defined as
\be
\left. P = 2n^2\lambda^2 h_y^2 - 2\mu^2 h =  2n^2\lambda^2 h_y^2 - 2\mu^2 h \right|_{y=y_p}
\ee
where the last equality follows from the fact that the pressure is constant in the BPS model. Hence
\be
P= 2n^2 \lambda^2 h_y(y_p)=2n^2\lambda^2 p^2 \;\;\; \Rightarrow \;\;\; p= - \frac{1}{n\lambda} \sqrt{\frac{P}{2}} .
\ee
Inserting this into (\ref{size-pres}) gives 
\be
y_p= \frac{2n\lambda}{\mu} \left[ \sqrt{1+\frac{P}{2\mu^2}} - \sqrt{\frac{P}{2\mu^2}} \right]
\ee
which leads to the right equation of state. 
%%%%%%%%%%%%%%%%%%%%%%%%%%%%%%%%%%%%%%%%%
\subsubsection{The gauged BPS baby Skyrme model}
%%%%%%%%%%%%%%%%%%%%%%%%%%%%%%%%%%%%%%%%%
Let us now apply this method for the gauged BPS Skyrme model with the assumption of a non-dynamical magnetic field. Then, the field equation leads to the general solution ($\beta = H/n$)
\be
h(y)=\frac{\mu^2}{2n^2\lambda^2\beta^2} \left[\ln (1+\beta y) + \frac{1+\beta y_0}{1+\beta y} \right] +C
\ee
where $y_0, C$ are integration constants. Again, the boundary conditions are 
\be
h(0)=1, \;\;\; h(y_p)=0, \;\;\; h_y(y_p)=p
\ee
where $y_p$ is the size of the compacton. Thus, the one $p$-parameter family of solutions reads
\be
h(y)=\frac{\mu^2}{2n^2\lambda^2 \beta^2} \left[ \ln (1+\beta y) - \frac{\beta y}{1+\beta y} (1+\beta y_0) \right] +1
\ee
where
\be
\frac{\mu^2}{2n^2\lambda^2 \beta^2} \left[ \ln (1+\beta y_p) - \frac{\beta y_p}{1+\beta y_p} (1+\beta y_0) \right] +1=0
\ee
and
\be
y_0=y_p-\frac{2n^2\lambda^2}{\mu^2} (1+\beta y_p)^2 p .
\ee
Again, the parameter $p$ must be related to the pressure by 
\be
\left. P=2n^2\lambda^2 (1+a)^2h_y \right|_{y=y_p}=\left. 2n^2\lambda^2 (1+\beta y )^2h_y \right|_{y=y_p} = 2n^2\lambda^2 (1+\beta y_p)^2 p^2
\ee
leading to
\be
p = - \frac{1}{n\lambda(1+\beta y_p)} \sqrt{\frac{P}{2}} .
\ee
Then, the relation between the size of the compacton $y_p$ and the pressure $P$ is
\be
\beta y_p \left( 1+ \beta \frac{2n\lambda}{\mu^2} \sqrt{\frac{P}{2}} \right) - \ln (1+\beta y_p) = \frac{2n^2\lambda^2}{\mu^2} \beta^2
\ee
which gives the following exact equation of state
\be \label{V.22}
\frac{HV}{2\pi n} \left( 1+ H \frac{2\lambda}{\mu^2} \sqrt{\frac{P}{2}} \right) - \ln \left(1+\frac{HV}{2\pi n} \right) = \frac{2\lambda^2}{\mu^2} H^2.
\ee
In Fig. 12 we plot the numerically determined equation of state for the full model together with the case without backreaction, for $g=0.2$. We find that both figures are quite similar. In Fig. 13, we plot the numerically determined equations of state for different values of $g$. 
Using (\ref{V.22}) it can be shown that the compressibility of the BPS
baby matter at any finite value of the external magnetic field is
still infinite
%%The compressibility of the BPS baby matter at any finite value of the external magnetic field is still infinite, 
\be
 \kappa=-\frac{1}{V} \left( \frac{\partial V}{\partial P} \right)_{H, \; P=0} = \infty .
\ee
We remark that this is a property of the {\em classical} field theory, which should be modified by quantum corrections. This will be relevant in applications where the quantization at least of some degrees of freedom is required, as, e.g., in applications to nuclear matter in three dimensions.
This is also the case for a non-zero value of the electromagnetic coupling.
As an example, we plot the numerical compressibility as a function of $P$ for $g=0.1$ and $H=0.3$ in Fig. 14.
\begin{figure}
%\hfill
\includegraphics[height=5.6cm]{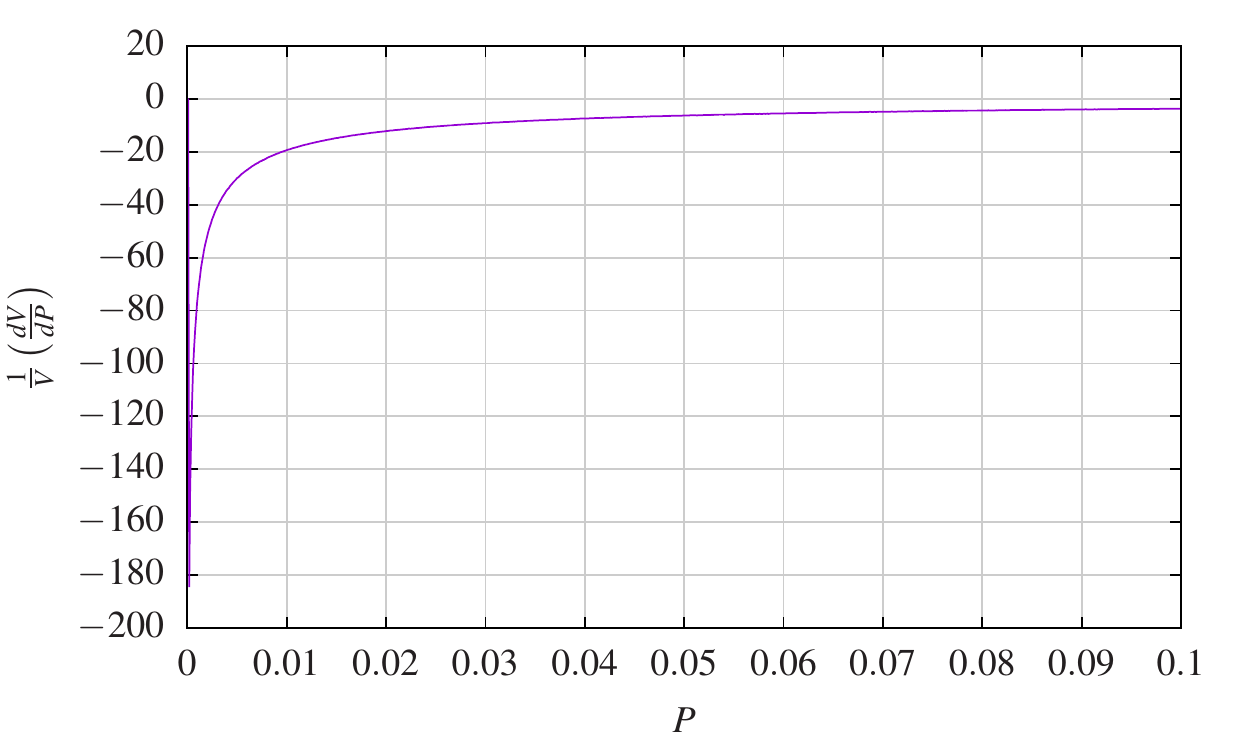}
% \hfill
\caption{The (isothermal) compressibility for different values of $g=0.1$ and $H=0.3$.}
\end{figure}

The magnetic compressibility is
\be
\kappa_{mag}^0(P)= \frac{1}{V} \left( \frac{\partial V}{\partial H} \right)_{P, \; H=0} = 
\frac{2\lambda}{3\mu} \left( \sqrt{1+\frac{P}{2\mu^2}} - \sqrt{\frac{P}{2\mu^2}} \right) 
\ee
which can be expressed in terms of the volume density at zero magnetic field
\be
\kappa_{mag}^0(P)=\frac{1}{6\pi n} V_0.
\ee
Here $V_0 \equiv V(H=0, P)$. 
The magnetic compressibility tends to zero as the pressure grows. This is an expected result. The higher pressure squeezes the compactons to smaller volumes (more dense matter). Hence, they behave stiffer under the action of the external magnetic field. 

\vspace*{0.2cm}

Further, the energy is
\be
E=-4\pi \frac{\mu^2n}{H} + \frac{\mu^4V}{2 \lambda^2H^2} \left[ \left( 1+\frac{HV}{2\pi n }\right)\left( 1+\frac{2\lambda H}{\mu^2} \sqrt{\frac{P}{2}} \right)^2 -1\right] 
\ee 
which together with the equation of state gives the $E=E(H,P)$ dependence. Notice that the field theoretical pressure still fulfills the thermodynamic relation
\be
\left( \frac{\partial E}{ \partial V} \right)_H = - P .
\ee
It is possible to express the energy as a function of two independent variables only. Namely, 
\be
E(H,V)=-4\pi \frac{\mu^2n}{H} + \frac{\mu^4V}{2 \lambda^2H^2} \left[ \left( 1+\frac{HV}{2\pi n }\right)\left[ \ln \left(  1+\frac{HV}{2\pi n } \right) +\frac{2\lambda^2n^2H^2}{\mu^2} \right]^2 \left(\frac{2\pi n}{HV} \right)^2-1\right] 
\ee
Then, at vanishing $H$ 
\be
E(H=0,V)=\frac{8\lambda^2n^2\pi^2}{V_0} + \mu^2 V_0 - \frac{\mu^4V_0^3}{96\lambda^2n^2\pi^2}
\ee
which reproduces the energy-pressure relation for the non-gauge case. Moreover, using the relation
\be
\left( \frac{\partial E}{\partial H} \right)_P = \left( \frac{\partial E}{\partial H} \right)_V +\left( \frac{\partial E}{\partial V} \right)_H \left( \frac{\partial V}{\partial H} \right)_P
\ee
we can find the magnetization density at $H=0$ as
\be
m^0(P)=-\frac{8 \pi \lambda^2 n}{3 }  \frac{1}{V_0} = - \frac{2 }{3} \frac{\lambda \mu}{\sqrt{1+\frac{P}{2\mu^2}} -\sqrt{\frac{P}{2\mu^2}}}
\ee
Hence, the negative magnetization of the medium is enhanced by the pressure. It is a consequence of the fact that the magnetization is pressure independent and therefore its density diverges for large $P$ as the volume shrinks. Another observation is that at $H=0$ the magnetic compressibility is proportional to the inverse of the magnetization density. Thus the following product is pressure independent
\be
m^0 \cdot \kappa_{mag}^0 = - \frac{4\lambda^2}{9} .
\ee
%%%%%%%%%%%%%%%%%%%%%%%%%%%%%%%%%%%%%%%%%
\section{An exact toy model}
%%%%%%%%%%%%%%%%%%%%%%%%%%%%%%%%%%%%%%%%%
Here we exploit the fact that BPS baby skyrmions exist even in the case without a potential, $U=0$, if the external pressure takes a non-zero value. Of course, in the limit $P \rightarrow 0$ the solitons disappear, in accordance with the Derrick theorem. Concretely, we show that for the model without potential one can find the equation of state $V=V(H,P)$ for any value of the coupling constant $g$. Indeed, now the BPS system (for any value of the pressure) may be solved exactly. 
%%%%%%%%%%%%%%%%%%%%%%%%%%%%%%%%%%%%%%%%%
\subsection{The BPS baby model}
%%%%%%%%%%%%%%%%%%%%%%%%%%%%%%%%%%%%%%%%%
The corresponding first order equation reads
\be
\frac{\lambda^2}{8}j_0^2=P \;\;\; \Rightarrow \;\;\; \lambda^2n^2h_y^2=P
\ee
where the axial ansatz together with the new target space and base space variables $h$ and $y$ has been used. The obvious solution is
\be
h=1-\frac{y}{y_0}, \;\;\; y_0=\frac{\sqrt{2}\lambda n}{\sqrt{P}}
\ee  
for $y \leq y_0$ and 0 otherwise. Then the equation of state is
\be
V^2P=8\pi^2 \lambda^2 n^2 .
\ee
%%%%%%%%%%%%%%%%%%%%%%%%%%%%%%%%%%%%%%%%%
\subsection{The gauged BPS baby model}
%%%%%%%%%%%%%%%%%%%%%%%%%%%%%%%%%%%%%%%%%
In the case without potential, the superpotential equation (in the case of external pressure) is
\be
\frac{\lambda^2}{4}W_h^2+g^2\lambda^4W^2=2P, \;\;\;\; W(0)=0, \;\; W^2_h(0)=\frac{8}{\lambda^2} P
\ee
It can be easily solved, 
\be
W=\frac{\sqrt{2P}}{g\lambda^2} \sin \left( 2g\lambda h  \right) .
\ee
Then, the BPS equations are
\be
2nh_y(1+a)=-\frac{1}{2}W_h = - \frac{\sqrt{2P}}{\lambda} \cos \left( 2g\lambda h  \right) ,
\ee
\be
na_y=-g^2\lambda^2W= - \sqrt{2P}g \sin \left( 2g\lambda h  \right) .
\ee
This can be further integrated to 
\be
\cos (2g\lambda h ) (1+a)=C_1
\ee
where $C_1$ is a constant. It is consistent with the first second order equation of motion for vanishing potential,
\be 
\partial_y [h_y(1+a)^2]=0, \;\; \Rightarrow \;\; h_y=\frac{C_1}{(1+a)^2} .
\ee
The constant $C_1$ can be found from the boundary condition at $y=0$. Indeed, $a(0)=0$ and $h(0)=1$ give $C_1=\cos (2g\lambda)$. Further, we can find a first order equation for the soliton profile
\be
2nh_y = -\frac{\sqrt{2P}}{\lambda C_1}\cos^2 (2g\lambda h)
\ee
with the solution
\be
\tan 2g\lambda h   = - \frac{g\sqrt{2P}}{nC_1} (y-C_2) \;\; \Rightarrow \;\; h = \frac{1}{2g\lambda} \arctan \frac{g\sqrt{2P}}{nC_1} (C_2-y)
\ee
The boundary conditions lead to 
\be 
C_2=y_0, \;\;\; \frac{C_2}{C_1}=\frac{n}{g\sqrt{2P}} \tan 2g\lambda
\ee
where $y_0$ denotes the compacton radius. Hence,
\be
h= \frac{1}{2g\lambda} \arctan\left[ \tan (2g\lambda) \left(1-\frac{y}{y_0} \right) \right] .
\ee
The equation for the magnetic field takes the following simple form
\be
(1+a)=\frac{C_1}{\cos (2g\lambda h)} = \cos (2g\lambda) \sqrt{1+ \tan^2(2g\lambda) \left(1-\frac{y}{y_0} \right)^2} .
\ee
So, finally, the gauge field has the following form
\be
a=-1+\cos (2g\lambda)\sqrt{1+ \tan^2(2g\lambda) \left(1-\frac{y}{y_0} \right)^2} 
\ee
which obeys $a(0)=0$ and $a_y(y_0)=0$. The asymptotic value is
\be
a_\infty = -1+\cos 2g\lambda .
\ee
Moreover, the size of the compacton is
\be
y_0=\frac{n\sin 2g\lambda}{g\sqrt{2P}} .
\ee
The corresponding equation of state is very similar to the non-gauge case
\be
V^2 P= \frac{2\pi^2}{g^2} n^2 \sin^2 2g\lambda .
\ee
%%%%%%%%%%%%%%%%%%%%%%%%%%%%%%%%%%%%%%%%%
\subsection{The gauged BPS baby model with asymptotically constant magnetic field}
%%%%%%%%%%%%%%%%%%%%%%%%%%%%%%%%%%%%%%%%%
It is convenient to use the "tilde" notation i.e., with the shifted superpotential 
\be
\frac{\lambda^2}{4}\tilde W_h^2+g^2\lambda^4\tilde{W}^2=\frac{H^2}{g^2}+ 2P\equiv 2\tilde P, \;\;\;\; \tilde W(0)=-\frac{H}{g^2\lambda^2} .
\ee
Now
\be
\tilde W=\frac{\sqrt{2\tilde P}}{g\lambda^2} \sin \left( 2g\lambda h +\beta  \right)
\ee
such that 
\be
\sqrt{2 \tilde P}\sin \beta = -\frac{H}{g} \;\; \Rightarrow \;\; \sin \beta = - \frac{H}{\sqrt{H^2+2 Pg^2}} .
\ee
Then, the BPS equations are
\be
2nh_y(1+a)=-\frac{1}{2}\tilde W_h = - \frac{\sqrt{2\tilde P}}{\lambda} \cos \left( 2g\lambda h  +\beta \right) ,
\ee
\be
na_y=-g^2\lambda^2 \tilde W= - \sqrt{2\tilde P}g \sin \left( 2g\lambda h +\beta \right) .
\ee
Repeating the same steps as before we find the following exact expression for the profile of the compactons
\be
\tan \left(  2g\lambda h + \beta \right) = \tan (2g\lambda +\beta) \left(1-\frac{g\sqrt{2\tilde P } \; y}{n \sin (2g\lambda +\beta)}  \right), \;\;\;\; y \leq y_0 
\ee
where 
\be
y_0=\frac{n\sin 2g\lambda}{g\sqrt{2\tilde P} \cos \beta}
\ee
and the corresponding solution for $a$,
\be
1+a=\frac{C_1}{\cos (2g\lambda h +\beta)} .
\ee
However, it is easy to show that
\be
\cos \beta =\frac{\sqrt{2P}}{\sqrt{2\tilde P}}
\ee
Then, the equation of state reads
\be
V^2P = \frac{2 \pi^2 n^2 \sin^2 2g\lambda}{g^2} ,
\ee 
which is exactly the same as in the usual gauge case. Hence, in contrast to the approximate but analytical results for the old baby potential, the asymptotically constant magnetic field 
does not change the size of the BPS baby skyrmions in the case without potential. Notice that the electromagnetic coupling constant does influence the equation of state, although the latter is $H$ independent. As a consequence, the BPS skyrmions for zero potential form a medium which is magnetically transparent.

All this shows that a particular form of the potential can drastically change the equation of state and some magnetic as well as thermodynamical properties of the BPS baby Skyrme matter. 
%%%%%%%%%%%%%%%%%%%%%%%%%%%%%%%%%%%%%%%%%
\section{Summary}
%%%%%%%%%%%%%%%%%%%%%%%%%%%%%%%%%%%%%%%%%
\noindent In the present paper we have continued the investigation of the gauged BPS baby Skyrme model.  One first main result is that the model exactly preserves its BPS property also for a nontrivial boundary condition for the magnetic field. In particular, it has been shown that, in the case of an asymptotically constant value of the magnetic field $B=H=const.$, there is a topological bound (for the regularized energy). Further the bound is saturated for configurations obeying BPS equations. If compared with the zero boundary value case ($H=0$), the BPS equations are modified additively by the inclusion of the boundary magnetic field. Moreover, also the superpotential equation slightly changes its form.  Both the BPS equations as well as the equation defining the superpotential may be brought to the former case ($H=0$) by a suitable redefinition of the target space variables and the potential. The information on the nontrivial asymptotical value of the magnetic field is then entirely encoded in the new boundary condition for the superpotential.  
\\
Moreover, using a recently proposed framework for the study of BPS models under non-zero external pressure \cite{pres}, we have shown how one can include pressure into the gauged BPS baby Skyrme model by a further, simple modification of the superpotential equation. 
\\
It is quite surprising that the different external parameters (pressure and external magnetic field) enter into the BPS equation in a very similar and in fact very natural manner. In addition, there is an intriguing similarity between the BPS equation with non-zero $H$ and $P$ and the non-extremal solutions in the fake supersymmetric theories \cite{susy}. 
\\
Another interesting observation is that certain global (integrated) quantities, like the (regularized) energy, the (compacton) volume, or the magnetization are, in fact, thermodynamic functions, i.e., they do not depend on the specific solution for which they are evaluated.  Instead, they give the same function of the external pressure $P$ and magnetic field $H$ for all equilibrium configurations (solutions of the BPS equations), and these thermodynamic functions obey the standard thermodynamic relations, like $M=(1/g^2)\Phi_{\rm reg}$ or 
$P=-(\partial E_{\rm reg}/\partial V)\vert_H$. Proving these relations is not trivial and requires the use of the BPS equations, so the standard thermodynamics of the theory is probably related to its BPS nature. 
More concretely, we proved the first relation, $M=(1/g^2)\Phi_{\rm reg}$, for zero pressure, but the generalization to nonzero pressure is trivial and just requires to replace the potential $U$ by the effective potential $U_{eff} = U+(P/\mu^2)$ in the proof. On the other hand, the second relation, $P=-(\partial E_{\rm reg}/\partial V)\vert_H$, has been proven only for the case without electromagnetic coupling in \cite{pres}, and for some specific examples in the present paper. The general proof should probably follow a strategy similar to the proof of the first relation in Section II.E, but is rendered more difficult due to the complicated expression (\ref{vol}) for the "volume" (area).

We emphasize again that because of the symmetries of the theory, the thermodynamic behaviour is completely independent of the shape of the skyrmions, and the model has the thermodynamic properties of a ferromagnetic perfect fluid.

The existence of baby skyrmions has been confirmed for the old baby potential. First of all, exact solutions have been found in the weak coupling regime, i.e., for the vanishing electrodynamic coupling constant $g$, which is equivalent to the non-back reaction limit. Then, the BPS equations can be solved analytically even with non-zero $H$ and $P$ leading not only to exact  solutions but, more importantly, to an exact equation of state, that is, a relation between the "volume" (area) and the pressure and external magnetic field at zero temperature, $V=V(P,H)$. Here the definition of the volume is straightforward, due to the compact nature of the baby skyrmions once the old baby potential is chosen. Further, the pressure, which is introduced in a standard field theoretic way as a component of the spatial part of the energy-momentum tensor, agrees with the {\it thermodynamical} pressure. For non-zero $g$, or for the system with dynamical gauge field, we performed numerical computations which, on the one hand, completely confirm the weak coupling approximation while, on the other hand, allow to understand the system also in the strongly coupled regime. Indeed, we have found the equation of state for any value of the electromagnetic coupling constant. Let us notice that the weak coupling approximation works surprisingly well even for quite big values of the coupling constant. Some quantities, as the susceptibility, are almost $g$ independent (for $H>0$). 

Qualitatively, turning on the external magnetic field has the following effects on the baby skyrmions.
\begin{itemize}
\item
The external magnetic field $H$ squeezes a baby skyrmion to a smaller size if has the same sign like the permanent magnetization $M$ of the skyrmion, while it enlarges the skyrmion if $H$ and $M$ have opposite signs. Concretely, for skyrmions with positive topological charge, where $M<0$, the external magnetic field squeezes skyrmions for $H<0$ and expands them for ($H>0$). For sufficiently large positive magnetic field we have observed a linear growth of the size of the solitons, while for $H \rightarrow - \infty$ the size decreases as $1/|H|$. 
\item If $H$ and $M$ have opposite signs and $H$ is sufficiently weak, then the phenomenon of magnetic flux inversion occurs. That is to say, the total magnetic field $B$ flips sign in a shell or skin region near the boundary of the skyrmion, because it has to take the value $B=H$ at the boundary. On the other hand, it preserves its original sign resulting from the permanent magnetization in the interior (core region) of the skyrmion. 
\item 
Both the magnetization of the skyrmion, $M=(1/g^2)\int d^2 x(B-H)$, and the magnetic susceptibility maintain their orientation (sign) for all values of the external magnetic field $H$ (negative for positive topological charge). The absolute value of the magnetization even grows for a large and oppositely oriented $H$, essentially because the skyrmion size grows. It goes, however, to a finite value in the limit $H\to \infty$, such that the magnetization density goes to zero in that limit. The same is true for the density of magnetic susceptibility.
\item
The main consequence of the equation of state is that the matter described by the gauged BPS Skyrme model behaves as a rather nonlinear ferromagnetic medium.  BPS baby skyrmions remain magnetized even when the external magnetic field vanishes, i.e., they possess a permanent magnetization. The magnetic properties of the BPS baby Skyrme matter may be made more pronounced by assuming sufficiently large values for the parameter $\lambda$. That is to say, depending on the values of the parameters, such a theory can model a weak as well as a strong magnetic medium.  
\item
As one might expect, the pressure always squeezes the solitons. Notice that the compressibility is always infinite for these classical skyrmion solutions, which already holds in the non-gauged model as a consequence of the quadratic approach to the vacuum for the old baby potential. This fact is not affected by gauging the model or by the external magnetic field. 
\end{itemize}

\vspace*{0.2cm}

There are many open questions and new directions in which the present work may be continued. \\
First of all, if we stay within the gauged BPS Skyrme model, there is the problem of the relation between a particular choice for the potential and the corresponding equation of state. If we restrict ourselves to the non-gauged case, then the analysis is very similar to the one performed recently in \cite{pres}. The volume-pressure relation can be easily found. Qualitatively different potentials are classified by their behavior near the vacuum (type of approach) leading to finite or infinite values of the compressibility. When we switch to the gauged version, the situation is more involved. In Ref. \cite{BPS-g}  it was found that there are no gauged solitons in the BPS model with double vacuum potentials (even in the non-BPS sector), which is in contrast with the non-gauged case, where BPS baby skyrmions do exist for potentials with both one or two vacua. However, as we observed in section VI, external pressure may allow for baby skyrmions even if such solutions disappear in the $P=0$ limit due to the Derrick theorem. Hence, it is reasonable to expect that, if a non-zero pressure is applied, skyrmions might appear also in the double vacuum potential case. Obviously, the resulting equation of state will have a singularity for $P=0$ or for some other (critical) values of the external parameters $P, H$. The necessary condition for the appearance of gauged baby skyrmions will be the existence of the superpotential $W$ (as a solution of the corresponding superpotential equation) on the whole interval $h \in [0,1]$.  As we now have two external parameters to play with, it should be possible by performing a fine tuning to find global solutions on the unit interval. This issue is under current investigation. 
\\
Secondly, it would be very interesting to check what happens if the Dirichlet term (the standard nonlinear sigma model term) is added to the energy. Such a modification of the gauged BPS model drastically changes its mathematical properties. The APD symmetries are explicitly broken (up to $U(1)$ rotations) and the theory is no longer BPS. It is also known that some crystal structures usually emerge \cite{td-baby1}. However, if we assume that the main contribution to the energy comes from the BPS part of the full model, i.e., the Dirichlet part is multiplied by a small parameter $\epsilon$, we are still in a near BPS regime with only softly broken APD symmetries (for a recent investigation of this issue, see \cite{sp2}). Hence, one may wonder whether, for sufficiently small $\epsilon$, we would continue to have liquid (plastic) ferromagnetic matter, as found for the BPS limit. Then, by increasing the value of $\epsilon$ (Dirichlet term) we could observe a transition into a crystal phase whose magnetic properties also remain to be found. 
For a phase diagram of the baby Skyrme model, but in a rather
different range of parameters, see \cite{td-baby2}.
\\
Unfortunately, the inclusion of the Dirichlet energy leads to several difficulties. As solitons become infinitely extended, one has to use an improved definition of the volume. However, there is an ambiguity in the definition of such a "physical" volume. Next, the pressure cannot be introduced by a BPS like equation, which, as a consequence, leads to the fact that it is not constant inside baby skyrmions. Nonetheless, the external pressure can still be introduced by a pertinent boundary condition representing solitons in a finite box (volume). Then, the field theoretical definition of the pressure would apply at the boundary. Combining these problems together we notice that now there is no reason for the field theoretical pressure to be also the thermodynamical pressure (that is to say, the thermodynamic relation $(\partial E/\partial V)_H = -P$ need no longer be true). This may result in a rather non-standard magnetothermodynamics.   
\\
Another straightforward generalization of the present research is to add the Chern-Simon term or to non-minimally couple the gauge potential to the topological current with the modification of the topological current to a gauge invariant (and still conserved) version \cite{schr1}. The main difference will be the appearance of a nontrivial temporal component of the gauge potential i.e., a nonzero electric field. Then, the APD symmetry of the energy integral will be lost. Since the (3+1) Skyrme model must also include the Wess-Zumino-Witten term, it is quite important to know how such a type of term can modify the equation of state and the magnetic properties of the medium. 
\\
In any case, as the baby Skyrme model found some applications in the context of condensed matter physics \cite{baby con}, it is natural to compare also its thermodynamical and magnetic properties with experimental data. It would be interesting to search for physical systems which might be described by the (BPS) baby Skyrme model and its thermodynamic properties, at least in a certain approximation.

\vspace*{0.2cm} 

Obviously, the most urgent issue is to perform an analogous analysis in the case of the BPS Skyrme model in (3+1) dimensions. The first step has already been done in \cite{pres}, where the thermodynamics at zero temperature has been investigated. The generalization to the gauged version (and its near BPS regime) is of high importance, as it would allow to understand the magnetic properties of BPS skyrmions and, therefore, some magnetic as well as thermodynamical properties of nuclear matter (for recent investigations of the magnetic properties of QCD see
\cite{mag}). 
%%%%%%%%%%%%%%%%%%%%%%%%%%%%%%%%%%%%%%%%%
\section*{Acknowledgement}
%%%%%%%%%%%%%%%%%%%%%%%%%%%%%%%%%%%%%%%%%
The authors acknowledge financial support from the Ministry of Education, Culture, and Sports, Spain (Grant No. FPA2008-01177), the Xunta de Galicia (Grant No. INCITE09.296.035PR and Conselleria de Educacion), the Spanish Consolider-Ingenio 2010 Programme CPAN (CSD2007-00042), and FEDER. Further, the authors acknowledge support from the Polish FOCUS grant (No. 42/F/AW/2014).  Further, A. W. was supported by the Polish NCN (National Science Center) Grant DEC-2011/01/B/ST2/ 00464 (2011-2014). The authors thank Wojtek Zakrzewski for his collaboration at an early stage of this research.

\end{document}